\documentclass[12pt]{JHEP3}
\usepackage{mathrsfs}
\usepackage{amsmath,amssymb}
\usepackage{psfrag}
\usepackage{relsize}
\usepackage{graphicx}

\def\x'{\mathaccent 19 x}
\def\y'{\mathaccent 19 y}
\def\n'{\mathaccent 19 n}
\def\u'{\mathaccent 19 u}

\def\et'{\mathaccent 19 \eta}
\def\th'{\mathaccent 19 \theta}
\def\lam'{\mathaccent 19 \lambda}
\def\varet'{\mathaccent 19 \vartheta}
\def\rh'{\mathaccent 19 \rho}
\def\ph'{\mathaccent 19 \phi}
\def\xb'{\mathaccent 19 {\bar{x}}}

\def\l{{\lambda}}

\def \A {{\bf{A}}}

\def\sl(2){\alg{sl}(2)}
\def \PP {{\cal P}}

\def\be{\begin{equation}}
\def\ee{\end{equation}}

\newcommand{\bea}{\begin{eqnarray}}
\newcommand{\eea}{\end{eqnarray}}
\newcommand\bp{\hbox{\larger\larger $\pi$}}
\newcommand{\eqn}[1]{(\ref{#1})}

\def\a {\alpha}
\def\b {\beta}
\def\s {\sigma}
\def\pa {\partial}

\def\g {\gamma}

\def\p{\phi}
\def\la{\label}

\def\ov{\over}

\def\str{\text{str}}

\def\G{\Gamma}
\def\S{\Sigma}

\newcommand{\alg}[1]{\mathfrak{#1}}
\newcommand{\su}{\alg{su}}

\newcommand{\psu}{\alg{psu}}

\newcommand{\AdS}{{\rm  AdS}_5\times {\rm S}^5}
\newcommand{\ads}{{\rm  AdS}_5\times {\rm S}^5}

\newcommand{\sfrac}[2]{{\textstyle\frac{#1}{#2}}}

\newcommand{\bem}{\left (\begin{matrix}}
\newcommand{\eem}{\end{matrix} \right )}

\newcommand{\nn}{\nonumber}

\newcommand{\Q}{\rm Q}
\newcommand{\sH}{\rm H}
\def\pws{p_{{\rm ws}}}

\author{Gleb Arutyunov$^a$\footnote{\smaller{Email:\mbox{
g.arutyunov@phys.uu.nl,~frolovs@maths.tcd.ie,~plefka@physik.hu-berlin.de,~marzam@aei.mpg.de}}}
{}\footnote{\smaller{Correspondent fellow at Steklov Mathematical
Institute, Moscow.}}\, , Sergey Frolov$^{b\, \dagger}$, Jan
Plefka$^{c}$ and Marija Zamaklar$^{d}$ \\  \\ $^{a}$
{\it \smaller{Institute for Theoretical Physics and Spinoza Institute,\\
~~Utrecht University, 3508 TD Utrecht, The Netherlands } }\\ \vspace{-.3cm}\\
$^b$ {\it \smaller{School of Mathematics, Trinity College, Dublin 2, Ireland}} \\
\vspace{-.3cm}\\
$^{c}$ {\it \smaller{Humboldt-Universit\"at zu Berlin, Institut
f\"ur Physik,\\ ~~Newtonstra{\ss}e 15, D-12489 Berlin, Germany}}\\\vspace{-.3cm}\\
$^{d}$ {\it \smaller{Max-Planck-Institut f\"ur Gravitationsphysik,
Albert-Einstein-Institut\\ ~~Am M\"uhlenberg 1, D-14476 Potsdam,
Germany}}} \abstract{ We analyze  the $\psu(2,2|4)$ supersymmetry
algebra of a superstring propagating in the $\ads$ background in
the uniform light-cone gauge. We consider the off-shell theory by
relaxing the level-matching condition and take the limit of
infinite light-cone momentum, which decompactifies the string
world-sheet. We focus on the  $\psu(2|2)\oplus\psu(2|2)$
subalgebra which leaves the light-cone Hamiltonian invariant and
show that it undergoes extension by a central element which is
expressed in terms of the level-matching operator. This result is
in agreement with the conjectured symmetry algebra of the dynamic
S-matrix in the dual ${\cal N}=4$ gauge theory. }

\title{The Off-shell Symmetry Algebra of the Light-cone \mathversion{bold}$\ads$\mathversion{normal} Superstring}

\preprint{\smaller{\smaller{\smaller{AEI-2006-071}}}\\[-.5ex]
          \smaller{\smaller{\smaller{HU-EP-06/31}}}\\[-.5ex]
          \smaller{\smaller{\smaller{ITP-UU-06-39}}}\\[-.5ex]
          \smaller{\smaller{\smaller{SPIN-06-33}}}\\[-.5ex]
          \smaller{\smaller{\smaller{TCDMATH 06-13}}}}

\begin{document}

\renewcommand{\thefootnote}{\arabic{footnote}}
\setcounter{footnote}{0}
%%%%%%%%%%%%%%%%%%%%%%%%%%%%%%%%%%%%%%%%%%%%%%%
\section{Introduction}
%%%%%%%%%%%%%%%%%%%%%%%%%%%%%%%%%%%%%%%%%%%%%%%

Understanding the symmetries of physical systems usually leads to the
most elegant way of solving them. The Green-Schwarz string theory on
the $\AdS$ background \cite{Metsaev:1998it} presents a prime example
of a system with a very large number of symmetries.  The manifest
global symmetry of the string sigma-model is given by the
$\text{PSU}(2,2|4)$ supergroup, an isometry group of the target coset
space. Moreover the world-sheet theory is a classically integrable
model \cite{Bena:2003wd}, possessing thus an infinite number of
(non)local integrals of motion. Finally, being a superstring theory,
it possesses the local gauge symmetries of world-sheet diffeomorphisms
and the fermionic $\kappa$-symmetry.

\medskip

Nevertheless, the quantisation of the superstring on $\AdS$ in a
covariant fashion is presently out of reach. However, fixing the
world-sheet gauge symmetry and reducing the theory to only
physical degrees of freedom simplifies the system and allows one
to invoke the notion of asymptotic states and the machinery of the
string S-matrix \cite{AFS,S,Beisert:2005tm}. Based on extensive
work \cite{Callan}-\cite{LCpaper} it is becoming more and more
apparent that a particularly convenient world-sheet gauge choice is
the so-called \emph{generalised uniform light-cone gauge}
\cite{AF,AFlc}. World-sheet excitations in this gauge are closely
related to the spin chain excitation picture appearing on the
gauge theory side, and thus make the link between the two more
transparent. Also, similarly as in flat space, this is the only
gauge which makes the Green-Schwarz fermions tractable.

\medskip

The generalised uniform light-cone gauge, however, has the unpleasant
feature that the gauge-fixed action does not possess world-sheet
Lorentz invariance. The fact that this symmetry is absent makes the
construction of the world-sheet S-matrix particularly involved:
standard constraints on the form of S-matrix arising from the
requirement of Lorentz invariance cannot be directly implemented, but
require subtle constructions \cite{Janik,AF06}. It is thus extremely
important to understand the symmetries of the world-sheet theory in
this gauge, how they constrain the form of the S-matrix, and how they
are connected with the symmetries present in the gauge theory.

\medskip

The generators of the superisometry algebra $\psu(2,2|4)$ of the
string sigma-model can be split into two groups: those which (Poisson)
commute with the Hamiltonian and those which do not. The former
comprise the $\su(2|2) \oplus \su(2|2)$ subalgebra of the full
$\psu(2,2|4)$ algebra, sharing the same central element which
corresponds to the Hamiltonian.  Another separation of the
$\psu(2,2|4)$ generators is into dynamical and kinematical generators,
depending on whether they do or do not depend on the unphysical field
$x_-$.

\medskip

There are three important facts related to the presence of the
unphysical field $x_-$ in the light-cone world-sheet theory. One
is that when solved in terms of physical fields, $x_-$ is a
\emph{non-local expression}.  Second is that the zero mode of the
field $x_-$ is a priori non-zero and has a non-trivial Poisson
bracket with the total light-cone momentum $P_+$. This implies
that dynamical charges change the light-cone momentum $P_+$.  It
also follows, that the zero mode part of the operator $e^{i \alpha
x_-}$ plays precisely the role of a ``length changing'' operator,
given that in the uniform light cone gauge, the total light cone 
momentum $P_+$ is
naturally identified with the circumference of the world-sheet
cylinder. However, and this shall be exploited extensively in the
present paper, in the limit of infinite light-cone momentum $P_+$,
the fluctuations of the $P_+$ are irrelevant, and the zero mode of
$x_-$ can be thus consistently ignored.
Thirdly, the differentiated field $x_-^\prime$ is a density of the
world-sheet momentum related to the presence of rigid symmetries
in the space-like $\sigma$-direction of the light-cone gauge fixed
string action. In the case of closed strings, the periodicity of
fields implies that the total world-sheet momentum $p_{\text{ws}}$ has to
vanish. This constraint is called the level-matching condition.

\medskip

In order to introduce the concept of the world-sheet excitations
as well as the world-sheet S-matrix, one needs to: (a) relax the
level-matching condition and (b) consider the limit $P_+
\rightarrow \infty$. If level-matching is not satisfied, we will
refer to the theory as ``off-shell''.  The limit (b) is necessary
in order to define asymptotic states and it basically defines the
gauge fixed world-sheet theory on the plane, rather than on a
cylinder of circumference $P_+$ \cite{MP}-\cite{magnon}. In this
paper, we will restrict the consideration to this limit, only
briefly commenting on the finite $P_+$ configurations in the
discussion.

\medskip

If the level-matching condition holds, the $\su(2|2)\oplus
\su(2|2)$ subalgebra of the $\psu(2,2|4)$ algebra is spanned by
those generators which commute with the world-sheet Hamiltonian.
Giving up the level-matching condition in string theory in
principle could spoil the on-shell $\psu(2,2|4)$ symmetry algebra
of the world-sheet theory and in particular the centrality of the
Hamiltonian with respect to the $\su(2|2)\oplus \su(2|2)$
subalgebra.  The main goal of this paper is the derivation of the
string world-sheet symmetry algebra in the case when the
level-matching is relaxed.

\medskip

In ${\cal N}=4$ gauge theory the analogue of the level-matching
condition is implemented by considering gauge-invariant operators,
i.e. traces of products of fields. Hence relaxing the
level-matching in string theory corresponds to ``opening'' the
trace of fields in gauge theory, i.e considering open rather
than closed spin chains.  Beisert has argued in
\cite{Beisert:2005tm} that opening the traces in gauge theory (and
considering infinitely long operators) leads to a modification of
the $\su(2|2)\oplus \su(2|2)$ algebra: the algebra receives two
central charges in addition to the Hamiltonian, which are
functions of the momentum carried by the one-particle excitations.
The Hamiltonian remains a central element.  This centrally
extended symmetry algebra beautifully allowed for the derivation
of the dispersion relation of the elementary ``magnon''
excitations as well as for the restriction of the form of S-matrix
down to one unknown function.

\medskip

The main result of this paper is the \emph{derivation} of the
centrally extended $\su(2|2)\oplus \su(2|2)$ algebra in 
\emph{string theory} both at the classical and quantum level. By
explicit computations, we show that relaxing the level-matching
condition in string theory in the limit of infinite light-cone
momentum necessarily leads to an enlargement of the
$\su(2|2)\oplus \su(2|2)$ algebra by a common central element
which is proportional to the level-matching condition. In
addition, the Hamiltonian remains central, as it was the case for
the on-shell algebra.

\medskip

The direct evaluation of the classical (and quantum)
$\su(2|2)\oplus\su(2|2)$ algebra is technically very difficult due to the
complexity of the supersymmetry generators and the non-canonical
Poisson structure of the theory. To derive the central charges, we
were thus forced to work in an approximation scheme which we named
the ``hybrid'' expansion scheme. Namely, in this approximation we expand
all supersymmetry generators in powers of fields (equivalently in
the inverse string tension $2\pi/\sqrt{\lambda}$),
keeping however all
dependence on the $x_-$ field \emph{intact} and  rigid. More precisely, the
dynamical charges depend on the $x_-$ field via $e^{i \alpha x_-}$ in a
multiplicative fashion. Although, when expressed in terms of the
physical fields this term is highly non-linear, in the ``hybrid''
expansion scheme we treat this quantity as a single object. This
allows us to determine \emph{the full, non-linear, bosonic part} of the
central charges. The fermionic part is then uniquely fixed from the
requirement that the central charges vanish on the level-matching
constraint surface.  Justification for the ``hybrid'' expansion scheme
is demonstrated in section 4.

%%%%%%%%%%%%%%%%%%%%%%%%%%%%%%%%%%%%%%%%%%%%%%%%%%%%%%%%%%%%%%%%%%%%

\section{Gauged-fixed string sigma-model}
In this section we collect the necessary background material
concerning the superstring theory on $\ads$. The central object on
which the construction of the string action is based on is the
well-known supersymmetry group ${\rm PSU(2,2|4)}$. We recall
\cite{Metsaev:1998it,Roiban:2000yy,Das:2004hy} that the
superstring action $\rm S$ is a sum of two terms: the (world-sheet
metric-dependent) kinetic term and the topological Wess-Zumino
term: \bea \label{sLag} {\rm S} =-{\sqrt{\lambda}\over
4\pi}\int_{-r}^{ r}\, {\rm d}\s{\rm d}\tau\,\Big(\gamma^{\a\b}{\rm
str}(\A^{(2)}_{\a}\A^{(2)}_{\b})+\kappa \epsilon^{\a\beta}{\rm
str}(\A^{(1)}_{\a}\A^{(3)}_{\beta}) \Big) \, , \eea Here
${\sqrt{\lambda}\over 2\pi}$ is the effective string tension,
coordinates $\s$ and $\tau$ parametrize the string world-sheet,
and for later convenience we choose the range of $\s$ to be $-
r\le \s\le  r$, where $r$ is an arbitrary constant. The standard
choice for a closed string is $r=\pi$. Then, $\gamma^{\a\b}
=\sqrt{-h}h^{\a\b}$ where $h^{\a\b}$ is the world-sheet metric,
and $\kappa=\pm 1$ to guarantee the invariance of the action
w.r.t. to the local $\kappa$-symmetry transformations. For the
sake of clarity we choose in the rest of the paper $\kappa=+1$.
Finally, $\A^{(i)}$ with $i=0,\ldots,3$ denote the corresponding
${\mathbb Z}_4$-projections of the flat current $\A=-g^{-1}{\rm
d}g$, where $g$ is a representative of the coset space
$$
\frac{\rm PSU(2,2|4)}{{\rm SO(4,1)}\times {\rm SO(5)}}\, .
$$
\medskip

The above-described Lagrangian formulation does not seem to be a
convenient starting point for studying many interesting properties of the theory, in particular,
for analyzing its symmetry algebra and developing a quantization procedure, because it suffers from the
presence of non-physical bosonic and fermionic degrees of freedom
related to reparametrization and $\kappa$-symmetry
transformations. A natural way to overcome this difficulty is to
use the Hamiltonian formulation of the model. It is obtained by
fixing the gauge symmetries and solving the Virasoro
constraints, the latter arise as equations of motion for the
world-sheet metric $h^{\a\beta}$.
Concerning the quantization, it should be implemented in such a way
that the global supersymmetry algebra, $\psu(2,2|4)$, which
includes the Hamiltonian, being restricted to physical states satisfying the level-matching condition would remain non-anomalous at the
quantum level. Hopefully, the quantum Hamiltonian could be
uniquely determined in this way and then the remaining problem
would be to determine its spectrum.
\medskip

A suitable gauge which leads to the removal of non-physical degrees of
freedom has been introduced in \cite{magnon}, following 
earlier work in \cite{Goddard:1973qh,Metsaev:2000yu,KT,AF,AFlc}.
We refer to it as the generalized uniform light-cone gauge. To
impose the generalized uniform light-cone gauge we make use of the
global AdS time, $t$, and an angle $\phi$ which parametrizes one
of the big circles of ${\rm S}^5$. They parametrize two ${\rm
U}(1)$ isometry directions of $\ads$, and the corresponding
conserved charges, the space-time energy $E$ and the angular
momentum $J$, are related to the momenta conjugated to $t$ and
$\p$ as follows \bea \nn E = - {\sqrt{\lambda}\over 2\pi} \int_{-
r}^{ r}\, {\rm d}\s\, p_t\ \, , \qquad J={\sqrt{\lambda}\over
2\pi} \int_{- r}^{ r}\, {\rm d}\s\, p_\p\ . \eea Then we introduce
the light-cone variables
$$
x_+=(1-a)t+a\phi\, , ~~~~~~~x_-=\phi-t
$$
whose definition involves one additional gauge parameter $a$:
$0\leq a\leq 1$. The corresponding canonical momenta are
$$
p_-=p_{t}+p_{\phi}\, ,  ~~~~~~~p_+=(1-a)p_{\phi}-ap_t\, .
$$
The reparametrization invariance is then used to fix the
generalized uniform light-cone gauge by requiring
that\footnote{Strictly speaking, $x_+$ can be identified with the
world-sheet time $\tau$ only for string configurations with
vanishing winding number. For the general consideration, see
\cite{AFlc, magnon}.} \bea\label{gauge} x_+=\tau\, ,
~~~~~~~p_+=1\, . \eea The consistency of this gauge choice
requires the constant $r$ to be \bea \la{Pplusdef} r
={\pi\ov\sqrt{\lambda}}P_+\equiv{1\over 2}\int_{- r}^{ r}\, {\rm
d}\s\, p_+\, , \eea where $P_+$ is the total light-cone
momentum.\footnote{In fact, $P_+$ has a conjugate variable
$x_-^{(0)}$ that is the zero mode of the light-cone coordinate
$x_-$. In any set of functions which do not contain $x_-$ the
variable $P_+$ plays the role of a central element and, therefore,
can be fixed to be a constant. }

Solving the Virasoro constraints, one is then left with 8 transverse coordinates $x_M$ and their conjugate momenta $p_M$.

The light-cone gauge should be supplemented by a suitable choice
of a $\kappa$-symmetry gauge which removes 16 out of 32
fermions from the supergroup element $g$  parametrizing the
coset $\frac{\rm PSU(2,2|4)}{{\rm SO(4,1)}\times {\rm
SO(5)}}$. The remaining 16 fermions $\chi$ have a highly
non-trivial Poisson structure which can be reduced to the canonical
one by performing an appropriate non-linear field redefinition, see
\cite{LCpaper} for detail.

\medskip

The gauged-fixed action that follows from (\ref{sLag}) upon fixing
(\ref{gauge}) and the $\kappa$-symmetry was obtained in
\cite{LCpaper}, c.f. also \cite{magnon}. Schematically, it has the
following structure \bea\la{Sgf} {\rm S}={\sqrt{\lambda}\over
2\pi} \int_{- r}^{ r}\, {\rm d}\s{\rm d}\tau\, \Big(p_M{\dot
x}_M+\chi^{\dagger}\dot{\chi}-{\cal H}\Big)\, , \eea where ${\cal
H}$ is the Hamiltonian density which is independent of both $\l$
and $P_+$, and is equal to $-p_-$. Since $p_-=p_{t}+p_{\phi}$, the
world-sheet Hamiltonian is given by the difference of the
space-time energy $E$ and the ${\rm U}(1)$ charge $J$:
\bea\label{GF} {\rm H}=-{\sqrt{\lambda}\over 2\pi} \int_{- r}^{
r}\, {\rm d}\s p_- = E-J\, . \eea

\medskip

Some comments are in order. As standard to the light-cone gauge
choice, the light-cone charge $P_+$ plays effectively the role of
the length of the string: in eq.(\ref{Sgf}) the dependence on
$P_+$ occurs in  the integration bounds only. Thus, the limit
$P_+\to\infty$ defines the theory on a plane and, by this reason,
it can be called the ``decompactifying limit''
\cite{MP}-\cite{magnon}. Obviously, for the theory on a plane one
should also specify the boundary conditions for physical fields.
As usual in soliton theory \cite{FadTaht}, several choices of the
boundary conditions are possible: the rapidly decreasing case, the
case of finite-density, etc. In what follows we will consider the
case of fields rapidly decreasing at infinity and show that it is
this case which leads to the realization of the centrally extended
$\su(2|2)\oplus \su(2|2)$ symmetry algebra.

\medskip

In the particular case $a=0$ we deal with the temporal gauge
$t=\tau$ analyzed in \cite{AF} which implies that the light-cone
charge $P_+$ coincides with the angular momentum $J$. In the rest
of the paper we will be mostly concentrated with the symmetric
choice $a=\frac{1}{2}$ studied in \cite{AFlc,LCpaper}. The reason
is that as was shown in those papers in this gauge the Poisson
structure of fermions simplifies drastically, and that makes it
easier to compute the Poisson algebra of the global symmetry
charges. All results we obtain, however, are also valid for the
general $a$ light-cone gauge.

\medskip

In what follows we find it convenient to use the inverse string
tension $\zeta=\frac{2\pi}{\sqrt{\lambda}}$, and to  rescale bosons $(p_M,x_M)\to
\sqrt{\zeta}(p_M,x_M)$ and fermions $\chi\to \sqrt{\zeta}\chi$
in order to ensure the canonical Poisson
brackets for the physical fields.
Upon these redefinitions the Hamiltonian (\ref{GF}) admits the following expansion in powers of $\zeta$
\bea\label{structureH} {\rm
H}=\int_{-r}^{r}{\rm d}\sigma \Big({\cal H}_2+\zeta {\cal
H}_4+\cdots\Big)\, . \eea
Here
the leading term ${\cal H}_2$ is quadratic in physical fields, and ${\cal
H}_4$ is quartic, and so on. Thus, $\zeta^{n-1}$ will be multiplied
by ${\cal H}_{2n}$ containing the product of $2n$ fields, and the expansion in $\zeta$ is an expansion
in the number of fields. The explicit expressions for ${\cal H}_2$ and ${\cal H}_4$
were derived in \cite{LCpaper} and we also present them in the Appendix 
to make the paper self-contained.

\medskip

To conclude this section we remark that the light-cone gauge does
not allow one to completely remove {\it all} unphysical degrees of
freedom. There is a non-linear constraint, known as the
level-matching condition, which is left unsolved. This constraint is just the statement that the total world-sheet momentum of the closed string vanishes, and it
reads in our case as\footnote{See Appendix for the notations.}
\bea \label{lm} {\pws}=-\zeta\int_{-r}^{r} {\rm
d}\sigma \Big(p_Mx'_M-\frac{i}{2}\str(\Sigma_+\chi\chi') \Big)=0\,
.
\eea
The variable $\pws$ generates rigid shifts in $\sigma$
and, therefore, in the limit $P_+\to\infty$ becomes a momentum
generator on the plane. We come to
the discussion of the influence of the level-matching constraint
on the supersymmetry algebra in the next section.

%%%%%%%%%%%%%%%%%%%%%%%%%%%%%%%
\section{Symmetry generators in the light-cone gauge}
\label{Sect:susygen} In this section we study the general
structure of the global symmetry generators in the light-cone
gauge and identify the subalgebra of symmetries of the gauge-fixed
Hamiltonian. We also reformulate a problem of computing the
Poisson brackets of symmetry generators in terms of the standard
notions of symplectic geometry.

\subsection{General structure of the symmetry
generators} The Lagrangian (\ref{sLag}) is invariant w.r.t. the
global action of the symmetry group ${\rm PSU(2,2|4)}$. The
generators of the Lie superalgebra $\psu(2,2|4)$ are realized by the
corresponding Noether charges which comprise\footnote{As explained
in \cite{AAT}, the part of $\rm Q$ which is proportional to the
identity matrix is not a generator of $\psu(2,2|4)$ and, therefore, it should be
factored out.} an $8\times 8$ supermatrix $\rm Q$. As was shown in
\cite{LCpaper}, in the light-cone gauge the matrix $\rm Q$ can be
schematically written as follows
%\footnote{We already implemented
%redefinitions of the previous section.}
\bea\la{charges} {\rm Q} =\int_{-r}^{r} {\rm d}\sigma~ \Lambda
U\Lambda^{-1}\, . ~~~~~ \eea An explicit form of the matrix $U$
can be found in \cite{LCpaper} and also in appendix 6.2, formula
(\ref{U}). It is important to note here that $U$ depends on
physical fields $(x,p,\chi)$ but not on $x_{\pm}$. The only
dependence of $\Q$ on $x_{\pm}$ occurs through the matrix
$\Lambda$ which is of the form \be \label{lambdadef}
\Lambda=e^{\frac{i}{2}x_+\Sigma_+ +\frac{i}{4}x_-\Sigma_-}\, ,\ee
where $\Sigma_{\pm}$ are the diagonal matrices of the form \bea
\Sigma_{\pm}={\rm diag}\Big(\pm 1,\pm 1, \mp 1,\mp 1;
1,1,-1,-1\Big)\, .\label{Sigmapm} \eea

We recall that the field $x_-$ is unphysical and can be solved in
terms of physical excitations through the equation
\bea\label{eqxmin} x'_-=-\zeta\Big( p_Mx'_M-\frac{i}{2}
\str(\Sigma_+\chi\chi') \Big)\, . \eea

Linear combinations of components of the matrix ${\rm Q}$
produce charges which generate rotations, dilatation,
supersymmetry and so on. To single them out one should multiply
${\rm Q}$ by a corresponding $8\times 8$ matrix ${\cal M}$, and
take the supertrace \bea \label{QM} {\rm Q}_{\cal M} = \str\,
({\rm Q}{\cal M}) \,. \eea In particular, the diagonal and
off-diagonal $4\times 4$ blocks of ${\cal M}$ single out bosonic
and fermionic charges of $\psu(2,2|4)$, respectively.

\medskip

Depending on the choice of ${\cal M}$ the charges ${\rm Q}_{\cal
M}\equiv {\rm Q}_{\cal M}(x_+,x_-)$ can be naturally classified
according to their dependence on $x_{\pm}$. Firstly, with respect
to $x_-$ they are divided into {\it kinematical} (independent of
$x_-$) and {\it dynamical} (dependent on $x_-$). Kinematical
generators do not receive quantum corrections, while the dynamical
generators do. Secondly, the charges, both kinematical and
dynamical, may or may not depend on $x_+=\tau$.

\medskip
In the Hamiltonian setting the conservation laws  have the
following form
$$
\frac{d \Q_{\cal M}}{d \tau}=\frac{\pa \Q_{\cal M}}{\pa
\tau}+\{\sH,\Q_{\cal M}\}=0\, .
$$
Therefore, the generators independent of $x_+=\tau$
Poisson-commute with the Hamiltonian. As follows from the Jacobi
identity, they must form an algebra which contains $\sH$ as the
central element.

\begin{figure}[t]
\begin{center}
\psfrag{M}{${\mathcal M} =$}
\psfrag{R}{Red}
\psfrag{B}{Blue}
\includegraphics*[width=.5\textwidth]{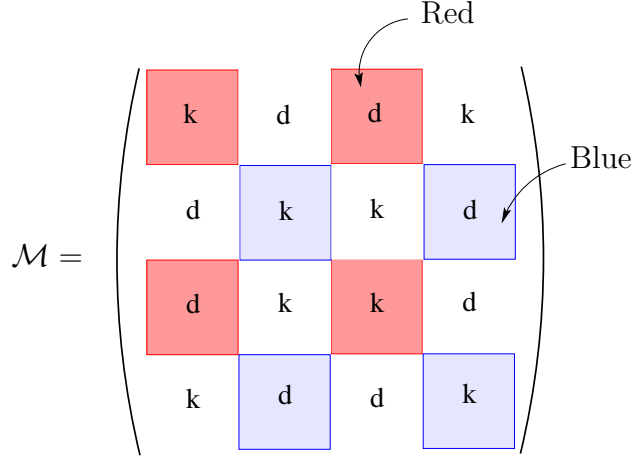}
\end{center}
\caption{\label{flag} The distribution of the kinematical and
dynamical charges in the ${\cal M}$ supermatrix. The red (dark) and blue (light)
blocks correspond to the subalgebra ${\cal J}$ of
$\psu(2,2|4)$ which leaves the Hamiltonian invariant.  }
\end{figure}

\medskip
Analyzing the structure of $\Q$ one can establish how a generic
matrix ${\cal M}$ is split into $2\times 2$ blocks each of them
giving rise either to kinematical or dynamical generators. This
splitting of ${\cal M}$ is shown in Figure 1, where the
kinematical blocks are denoted by ${\bf k}$ and the dynamical ones
by $\bf d$ respectively. Furthermore, one can see that the blocks
which are colored in red and blue give rise to charges which are
independent of $x_+=\tau$; by this reason these charges commute
with the Hamiltonian. Complementary, we note that the uncolored
both kinematical (fermionic)  and dynamical (bosonic) generators
do depend on $x_+$.

\medskip

These conclusions about the structure of ${\cal M}$ can be easily
drawn by noting that $\Lambda$ in eq.(\ref{lambdadef}) is built
out of two commuting matrices $\Sigma_{+}$ and $\Sigma_-$ (see
\eqn{Sigmapm}). For instance, leaving in ${\cal M}$ the kinematical
blocks only, i.e. ${\cal M}\equiv {\cal M}_{\rm kin}$, we find
that $[\Sigma_-, {\cal M}_{\rm kin}]$=0 and, therefore, due to the
structure of $\Q_{\cal M}$, see eq.(\ref{QM}), the variable $x_-$
cancels out in $\Q_{\cal M}$. Explicitly one finds the following
conjugation expressions with $\Lambda$ of \eqn{lambdadef}
\begin{align}
\Lambda^{-1}\, {\cal M}_{\rm dyn}^{\rm odd}\, \Lambda & = e^{
-\frac{i}{2} x_- \Sigma_-}\,
 {\cal M}_{\rm dyn}^{\rm odd} \, ,\qquad
&\Lambda^{-1}\, {\cal M}_{\rm dyn}^{\rm even}\, \Lambda & = \Lambda^2\,  {\cal M}_{\rm dyn}^{\rm even} \, , \nn \\
\Lambda^{-1}\, {\cal M}_{\rm kin}^{\rm odd}\, \Lambda & = e^{i\,
x_+\, \Sigma_+}\,  {\cal M}_{\rm kin}^{\rm odd} \, ,\qquad
&\Lambda^{-1}\, {\cal M}_{\rm kin}^{\rm even}\, \Lambda & = {\cal
M}_{\rm kin}^{\rm even}  \nonumber
\end{align}
showing that the $x_+=\tau$ independent matrices are indeed given
by ${\cal M}_{\rm dyn}^{\rm odd}$ and ${\cal M}_{\rm kin}^{\rm
even}$, i.e. by the red and blue entries in Figure \ref{flag}.

\medskip

Finally, we note that the Hamiltonian itself can be obtained from
$\Q$ as follows \bea {\rm H} =- {i\ov 2}\str\, (\Q\S_+)\,.
\label{Ham} \eea Another integral of motion, $P_+$, is given by
\begin{equation}
\label{HH} P_+=\frac{i}{4}\str(\Q\S_-)\, . \
\end{equation}
The structure of $\Q$ discussed above is found for finite $r$ and it also remains
valid in the limit $r\to \infty$.

%%%%%%%%%%%%%%%%%%%%%%%%%%%%%%%%%%%%%%%%%%%%%%%%%%%%
\subsection{Moment map and the Poisson brackets}
The group ${\rm PSU(2,2|4)}$ acts on the coset space $\frac{\rm
PSU(2,2|4)}{{\rm SO(4,1)}\times {\rm SO(5)}}$ by multiplications
of a coset element by an element of the group from the left. When
we fix the light-cone gauge and solve the Virasoro constraints we
obtain the well-defined symplectic structure $\omega$ (the inverse
of the Poisson bracket) for physical fields. Therefore, we are now
able to study the Poisson algebra of the Noether charges
corresponding to the infinitesimal global symmetry transformations
generated by the Lie algebra $\psu(2,2|4)$. Primarily we are
interested in those charges which leave the gauge-fixed Hamiltonian and, as the consequence, the
symplectic structure of the theory invariant; the corresponding
subspace in $\psu(2,2|4)$ will be called ${\cal J}$.

\medskip

Since the symplectic form $\omega$ remains invariant under the
action of $\cal J$, to every element $\cal M\in {\cal J}$ one can
associate a locally Hamiltonian phase flow $\xi_{\cal M}$ whose
Hamiltonian function is nothing else but the Noether charge
$\Q_{\cal M}$: \bea \omega(\xi_{\cal M},\ldots)+{\rm d}\Q_{\cal
M}=0\, . \label{HE} \eea Identifying  $\psu(2,2|4)$ with its dual
space, $\psu(2,2|4)^*$, by using the supertrace operation, we can
treat the matrix $\Q$ as the {\it moment map} \cite{Souriau} which
maps the phase space $(x,p,\chi)$ into the dual space to the Lie
algebra:
$$
{\rm Q}:~~~(x,p,\chi)\to \psu(2,2|4)^*\,
$$
and it allows one to associate to any element of $\psu(2,2|4)$ a
function ${\rm Q}_{\cal M}$ on the phase space. This linear
mapping from the Lie algebra into the space of functions on the
phase space is given by eq.(\ref{QM}). The function $\Q_{\cal M}$
appears to be a Hamiltonian function, i.e. it obeys eq.(\ref{HE}), only if ${\cal M}\in
{\cal J}$. Although the elements of $\psu(2,2|4)$ which do not belong to
$\cal J$ are the symmetries of the gauge-fixed action, they leave
invariant neither the Hamiltonian nor the symplectic structure.

\medskip

As is well known \cite{Kirillov,Arnold}, eq.(\ref{HE}) implies the
following general formula for the Poisson bracket of the Noether
charges $\Q_{\cal M}$ \bea \label{PBG} \{{\rm Q}_{{\cal M}_1},
{\rm Q}_{{\cal M}_2} \}=(-1)^{\pi({\cal M}_1)\pi({\cal
M}_2)}\str({\rm Q}[{\cal M}_1, {\cal M}_2])\, + {\rm C}({\cal
M}_1,{\cal M}_2) \, ,  \eea where ${\cal M}_{1,2}\in {\cal J}$.
Here $\pi$ is the parity of a supermatrix and $[{\cal M}_1, {\cal
M}_2]$ is the graded commutator, i.e. it is the anti-commutator if
both ${\cal M}_1$ and ${\cal M}_2$ are odd matrices, and the
commutator if at least one of them is even. The first term in the
r.h.s. of eq.(\ref{PBG}) reflects the fact that the Poisson
bracket of the Noether charges $\Q_{{\cal M}_1}$ and $\Q_{{\cal
M}_2}$ gives a charge corresponding to the commutator $[{\cal
M}_1,{\cal M}_2]$. The normalization prefactor $(-1)^{\pi({\cal
M}_1)\pi({\cal M}_2)}$ is of no great importance and, as we will
see later on, it is related to our specific choice of normalizing
the even elements with respect to the odd ones inside the matrix
$\Q$. The quantity ${\rm C}({\cal M}_1,{\cal M}_2)$ in the r.h.s.
of eq.(\ref{PBG}) is the central extension, i.e. a bilinear graded
skew-symmetric form on the Lie algebra ${\cal J}$ which
Poisson-commutes with all $\Q_{\cal M}$, ${\cal M}\in {\cal J}$.
The Jacobi identity for the bracket (\ref{PBG}) implies that ${\rm
C}({\cal M}_1,{\cal M}_2)$ is a two-dimensional cocycle of the Lie
algebra $\cal J$. For simple Lie algebras such a cocycle
necessarily vanishes, while for super Lie algebras it is generally
not the case. Since we consider a finite-dimensional super Lie
algebra  the central extension vanishes if the element $\cal M$ is
bosonic:  ${\rm C}({\cal M},\ldots)=0$.

\medskip

Some comments are necessary here. As we already mentioned in section 2, the usual feature
of closed string theory considered in the light-cone gauge is the
presence of the level-matching constraint $p_{\rm ws}=0$. This
constraint arises from the requirement of the unphysical
field $x_-$ to be a periodic function of the world-sheet coordinate
$\sigma$. The level-matching constraint cannot be solved in
classical theory, rather it is required to vanish on physical
states. Thus, before we turn our attention to the question of
the physical spectrum we should treat $p_{\rm ws}$ as a non-trivial
variable. We will refer to a theory with a non-vanishing generator $p_{\rm ws}$ as
the {\it off-shell} theory. Since the Hamiltonian contains
only physical fields it commutes with $p_{\rm ws}$: $\{{\rm H}, p_{\rm ws}\}=0$,
i.e. the momentum $p_{\rm ws}$ is an integral of motion.
The Poisson bracket (\ref{PBG}) with the vanishing central term is valid
on-shell and it is the off-shell theory where one could expect
the appearance of a non-trivial central extension.
Below we determine a general form of the central extension
based on symmetry arguments only and in the next section by explicit evaluation of the Poisson brackets
we justify the formula (\ref{PBG}) and also find a concrete realization of ${\rm C}({\cal M}_1,{\cal M}_2)$.

\medskip

Let us note that a formula as (\ref{PBG}) makes it easy to reobtain
our results on the structure of ${\cal J}$. Indeed, from
eq.(\ref{PBG}) we find that the invariance subalgebra ${\cal
J}\subset \psu(2,2|4)$ of the Hamiltonian is determined by the
condition
$$
\{{\rm H},\Q_{\cal M}\}=\str(\Q[\Sigma_+,{\cal M}])=0\, .
$$
Thus, $\cal J$ is the stabilizer of the element $\Sigma_+$ in $\psu(2,2|4)$:
$$
[\Sigma_+,{\cal M}]=0\, , ~~~~~~{\cal M}\in {\cal J}\, .
$$
Obviously, ${\cal J}$ coincides with the red-blue submatrix of
${\cal M}$ in Figure 1. Thus, for $P_+$ being finite\footnote{As
a side remark, we note that for $P_+$ finite the subalgebra
which leaves invariant both ${\rm H}$ and $P_+$ coincides with the
even (bosonic) sublagebra ${\cal J}_{\rm even}$ of $\cal J$.
According to eqs.(\ref{Ham}) and (\ref{HH}), this subalgebra
arises as the simultaneous solution the two equations,
$[\Sigma_+,{\cal M}]=0$ and $[\Sigma_-,{\cal M}]=0$, and it is
represented by the red and blue diagonal blocks in Figure 1.} we
would obtain the following vector space decomposition of ${\cal J
}$
$$
{\cal J}=\psu(2|2)\oplus\psu(2|2)\oplus \Sigma_+\oplus \Sigma_-\,
.
$$
The rank of the latter subalgebra is six and it coincides with
that of $\psu(2,2|4)$. In the case of infinite $P_+$ the last
generator decouples.

\medskip

Now we are ready to determine the general form of the central term in
eq.(\ref{PBG}). Denote by ${\cal J}_{\rm even}\subset {\cal J}$
the even (bosonic) subalgebra of $\cal J$. It is represented by
the red and blue diagonal blocks in Figure 1. Let $G_{\rm even}$
be the corresponding group. The adjoint action of $G_{\rm even}$
preserves the ${\mathbb Z}_2$-grading of $\cal J$. Obviously, if
we perform the transformation
$$
{\rm Q}\to g{\rm Q}g^{-1}\, , ~~~~~~{\cal M}\to g^{-1}{\cal M}g
$$
with an element $g\in G_{\rm even}$ the charge $\Q_{\cal M}$
remains invariant. This transformation leaves the l.h.s of the
bracket (\ref{PBG}) invariant. Thus, the central term must satisfy
the following invariance condition: \bea\label{invar} {\rm
C}(g{\cal M}_1g^{-1},g{\cal M}_2g^{-1})={\rm C}({\cal M}_1,{\cal
M}_2)\, . \eea It is not difficult to find a general expression
for a bilinear graded skew-symmetric form on ${\cal J}$ which
satisfies this condition. It is given by \bea\label{cocycle} {\rm
C}({\cal M}_1, {\cal M}_2) = \str \bigg(\left(\varrho {\cal M}_1
\varrho {\cal M}_2^t + (-1)^{\pi({\cal M}_1) \pi({\cal M}_2)}
\varrho{\cal M}_2 \varrho {\cal M}_1^t \right) \Delta  \bigg)\, .
\eea Here
\begin{equation}
\label{Delta} \Delta = - \frac{1}{2} {\footnotesize\left(
\begin{array}{cccc}
        c_3 \, {\mathbb I}_2 & 0 & 0 & 0 \\
        0 & c_1 \, {\mathbb I}_2 & 0 & 0 \\
        0 & 0 &  c_4 \, {\mathbb I}_2 & 0 \\
        0 & 0 & 0 & c_2 \, {\mathbb I}_2
\end{array} \right)}\, ,
\end{equation}
where ${\mathbb I}_2$ is the two-dimensional identity matrix and
\begin{equation}
\varrho =  {\footnotesize\left( \begin{array}{cccc}
        \sigma & 0 & 0 & 0 \\
        0 & \sigma & 0 & 0 \\
        0 & 0 & \sigma & 0 \\
        0 & 0 & 0 & \sigma
\end{array} \right)}\, , \quad \quad
\sigma ={\footnotesize\left( \begin{array}{cc}
        0 & 1 \\
        -1 & 0  \\
      \end{array} \right)}\, .
\end{equation}
Condition (\ref{invar}) follows from the form of the matrix
$\Delta$ and the equation
\begin{equation}
{\mathcal J}_{\rm even}^t \varrho + \varrho{\mathcal J}_{\rm even}
= 0 \, .
\end{equation}
The coefficients $c_i$, $i=1,\ldots, 4$ can depend on the physical
fields and they are central w.r.t. the action of $\cal J$:
$$
\{c_i,{\rm Q}_{\cal M}\}=0\, , ~~~~~{\cal M}\in {\cal J}\, .
$$
By using eq.(\ref{cocycle}) one can check that the cocycle
condition for ${\rm C}({\cal M}_1,{\cal M}_2)$ is trivially
satisfied. In accordance with our assumptions, ${\rm C}({\cal
M}_1,{\cal M}_2)$ does not vanish only if both ${\cal M}_1$ and
${\cal M}_2$ are odd.

\medskip

Taking into account that $\cal J$ contains two identical
subalgebras $\psu(2|2)$ we can put $c_1=c_3$ and $c_2=c_4$. Thus,
general symmetry arguments fix the form of the central extension
up to two central functions $c_1$ and $c_2$.  Since we consider
the algebra $\psu(2,2)$, which is the real form of $\alg{
psl}(2|2)$, the conjugation rule implies that $c_1= -c_2^*$. In the
next section we compute the Poisson brackets of the Noether
charges and determine the explicit form of $c\equiv c_1$.

%%%%%%%%%%%%%%%%%%%%%%%%%%%%%%%%%%%%%%%%%%%%%%%%%%%%%%%%%
\subsection{Explicit basis}
In what follows we find it convenient to pick up a basis in the space of the
Noether charges $\Q_{\cal M}$ with ${\cal M}\in {\cal J}$ and write down the bracket (\ref{PBG})
for the corresponding basis elements.
Since $\cal J$ contains two identical $\psu(2,2)$ sublagebras it
is enough to analyze the Poisson brackets corresponding to one of them.
For definiteness, we concentrate our attention on the subalgebra $\psu(2,2)_R$
which is represented in Figure 1 by blue (right) blocks.
For this subalgebra we select  a basis in
which the fermionic (dynamical) generators are given by \bea \label{BcB} Q^1_1 &=& {1\ov
2}\str\,{\rm Q}\sigma^+\otimes\left(
\G_{14}-\G_{23}+\PP_-\right)\,,~~~~~~~~~Q^1_2 =  \str\,{\rm Q}\sigma^+\otimes\G_{13}\,,\nonumber \\
Q^2_2 &=&  -{1\ov 2}\str\,{\rm Q}\sigma^+\otimes\left(
\G_{14}-\G_{23}-\PP_-\right) \,,~~~~~~~
Q^2_1
= -\str\,{\rm Q}\sigma^+\otimes\G_{42}\,  \eea and their conjugate charges
$\bar{Q}_a^\alpha$ are
\begin{eqnarray}
\label{Bcc} \bar{Q}^1_1 &=& - {1\ov 2}\str\,{\rm Q}\sigma^-\otimes\left(
\G_{14}-\G_{23}+\PP_-\right)\,,~~~~~~~~~\bar{Q}^1_2 =  - \str\,{\rm Q}\sigma^-\otimes\G_{13}\,, \nonumber \\
\bar{Q}^2_2 &=&  {1\ov 2}\str\,{\rm Q}\sigma^-\otimes\left(
\G_{14}-\G_{23}-\PP_-\right) \,, ~~~~~~~~~~~~
\bar{Q}^2_1 =   \str\,{\rm Q}\sigma^-\otimes\G_{42}\, .
\end{eqnarray}
On the other hand, the bosonic (kinematical) charges are defined as  \bea
\label{BL} L^1_1 &=& {i\ov 2}\str\,{\rm Q} P_2^+\otimes\left(
\G_{14}-\G_{23}\right) \,,~~~\quad L_2^2 = -L^1_1\,, \nonumber \\
L^1_2 &=&-i\, \str\,{\rm Q} P_2^+\otimes\G_{13}\,, ~~~~~~~~~~~~\quad L^2_1 = -i\,
\str\,{\rm Q} P_2^+\otimes\G_{24}\,, \eea and \bea \label{BR}
R^1_1 &=& {i\ov 2}\str\,{\rm Q} P_2^-\otimes\left(
\G_{14}-\G_{23}\right) \,,~~~~~\quad R_2^2 = -R^1_1\,, \nonumber \\
R^1_2 &=& i\, \str\,{\rm Q} P_2^-\otimes\G_{13}\,, ~~~~~~~~~~~~~~~~~\quad R^2_1 = i\,
\str\,{\rm Q} P_2^-\otimes\G_{24}\, , \eea We refer the
reader to the appendix \ref{appA}. for notations used here.

\medskip

Rewriting the bracket (\ref{PBG}) in this basis we obtain the following relations
\bea \label{Beis-alg}
\{R^a_b,J^c\}&=&\delta_b^c J^a - {1\ov 2}\delta_b^a J^c\,,\qquad
\{L^\a_\b,J^\g\}=\delta_\b^\g J^\a - {1\ov 2}\delta_\b^\a J^\g\,, \nonumber \\
\{ Q^\a_a, \bar{Q}^b_\b\} &=& \delta_a^b L^\a_\b + \delta^\a_\b R_a^b +{1\ov 2}\delta_a^b\delta^\a_\b {\rm H}\,, \nonumber \\
\{ Q^\a_a, Q^\b_b\} &=& \epsilon^{\a\b}\epsilon_{ab}~c\, ,
~~~~~~~~ \{ \bar{Q}_\a^a, \bar{Q}_\b^b\} =
\epsilon^{ab}\epsilon_{\a\b}~c^* \,. \eea Here in the first line
we have indicated how the indices $a$ and $\gamma$ of any Lie
algebra generator transform under the action of the bosonic
subalgebras generated by $R^a_b$ and $L^{\a}_{\b}$. In the next
section we are going to derive the so far undetermined central
function $c$ in terms of physical variables of string theory.

%%%%%%%%%%%%%%%%%%%%%%%%%%%%%%%%%%%%%%%%%%%%%%%%%%%%%%%%%%%%%%%%%%%

\renewcommand{\thefootnote}{\arabic{footnote}}
\setcounter{footnote}{0}
%%%%%%%%%%%%%%%%%%%%%%%%
\section{Deriving the central charges}

\def\z{\zeta}

%%%%%%%%%%%%%%%%%%%%%%%%\subsection{Hybrid expansion}

Given the complex structure of the supersymmetry generators in the
light-cone gauge as well as the corresponding Poisson structure of
the theory, the direct computation of the classical and quantum
supersymmetry algebra does not seem to be feasible. Hence,
simplifying perturbative methods need to be applied. The
perturbative expansion of the supersymmetry generators in powers
of $\zeta=\frac{2\pi}{\sqrt{\lambda}}$ or, equivalently, in the
number of fields defines a particular expansion scheme. This
expansion, however, does not allow one to determine the exact form
of the central charges because they are also expected to be non-trivial functions of
$\zeta$.
% as one can see from
%(\ref{Beis-alg}) and (\ref{lm}), they also admit an expansion in
%$\zeta$.
To overcome this difficulty  in this section we describe
a ``hybrid'' expansion scheme which can be used to determine the
exact form of the central charges. To be precise we determine only
the part of the central charges which is independent of fermionic
fields. We find that this part depends solely on the piece of
the level-matching constraint which involves the bosonic fields. Since the central charges must
vanish on the level-matching constraint surface, the exact form of the
central charges is, therefore, unambiguously fixed by its bosonic
part.

\medskip

More precisely, a dynamical supersymmetry generator has the
following generic structure
%\footnote{A kinematic symmetry
%generator has no $x_-$-dependence.}
\bea {\rm Q}_{\cal M}=\int {\rm d}\s
~e^{i\a x_-} \Omega( x,p,\chi;\zeta)\,. \label{sch} \eea Depending
on the supercharge, the parameter $\a$ in the exponent of
(\ref{sch}) can take two values $\a=\pm \frac{1}{2}$. Here the
function $\Omega(x,p,\chi;\zeta)$ is a {\it local} function of
transverse bosonic fields and fermionic variables. It depends on
$\zeta$, and can be expanded, quite analogous to the Hamiltonian,
in power series
$$ \Omega(x,p,\chi;\zeta)=
\Omega_{2}(x,p,\chi)+\zeta \Omega_{4}(x,p,\chi)+\cdots
$$
Here $\Omega_{2}(x,p,\chi)$ is quadratic in fields,
$\Omega_{4}(x,p,\chi)$ is quartic and so on. Clearly, every term
in this series also admits a finite expansion in number of
fermions. In the usual perturbative expansion we would also have
to expand the non-local ``vertex'' $e^{i\a x_-}$ in powers of
$\zeta$ because $x'_- \sim -\zeta px' +\cdots$. In the hybrid
expansion we do not expand $e^{i\a x_-}$ but rather treat it as a
rigid object.

\medskip

The complete expression for a supercharge is rather cumbersome.
However, we see that the supercharges and their algebra can be
studied perturbatively: first by expanding up to the given order
in $\zeta$ and then by truncating the resulting polynomial up to
the given number of fermionic variables. Then, as was discussed
above the exact form of the central charges is completely fixed by
their parts which depend only on bosons. Thus, to determine these
charges it is sufficient to consider the terms in $\Q_{\cal M}$ which are
linear in fermions, and compute their Poisson brackets (or
anticommutators in quantum theory) keeping only terms independent
of fermions. This is, however, a complicated problem because the
Poisson brackets of fermions appearing in (\ref{charges}) have a
highly non-trivial dependence on bosons, see \cite{LCpaper} for
details. It was shown in \cite{LCpaper} that to have the canonical
Poisson brackets one should perform a field redefinition which can
be determined up to any given order in $\zeta$. Taking into
account the field redefinition, integrating by parts if necessary,
and using the relation $x'_- \sim -\zeta px' +\cdots$, one can
cast any supercharge (\ref{sch}) to the following symbolic form
\bea {\rm Q}_{\cal M}=\int {\rm d}\s ~e^{i\a x_-}\,\chi\cdot\big(
\Upsilon_1(x,p) + \z \Upsilon_3(x,p) + \cdots \big) + {\cal O}(\chi^3)\,,
\label{sch2} \eea where $\Upsilon_1$ and $\Upsilon_3$ are linear
and cubic in bosonic fields, respectively. The explicit form of
the supercharges expanded up to the order $\z$ can be found in
the Appendix.

\medskip

It is clear now that the bosonic part of the Poisson bracket of
two supercharges is of the form
\bea\label{pbsch} \{{\rm Q}_1,
{\rm Q}_2\}\sim \int_{-\infty}^\infty {\rm d}\s ~&& e^{i(\a_1+\a_2)
x_-}\Big(
\Upsilon_1^{(1)}(x,p) \Upsilon_1^{(2)}(x,p)\\
 &&+ \z\big( \Upsilon_1^{(1)}(x,p) \Upsilon_3^{(2)}(x,p)
+ \Upsilon_3^{(1)}(x,p) \Upsilon_1^{(2)}(x,p)\big)+  \cdots \Big)\,, \nonumber
\eea
where $\Q_{1,2}\equiv \Q_{{\cal M}_{1,2}}$.
Computing the product $\Upsilon_1^{(1)}(x,p)
\Upsilon_1^{(2)}(x,p)$ in the case $\a_1=\a_2 =\pm 1/2$, we find
that it is given by
\bea\label{uusch} \Upsilon_1^{(1)}(x,p)
\Upsilon_1^{(2)}(x,p)\sim {1\ov \z}x_-' + {d\ov d\s}f(x,p)\,,
\eea
where $f(x,p)$ is a local function of transverse coordinates and
momenta. The first term in (\ref{uusch}) nicely combines with
$e^{\pm i x_-}$ to give ${d\ov d\s}e^{\pm i x_-}$, and integrating
this expression over $\s$, we obtain the sought for central charges
\bea\la{Ccg}
\int_{-\infty}^\infty {\rm d}\s ~{d\ov d\s}e^{\pm i x_-} =
e^{\pm i x_-(\infty)} - e^{\pm i x_-(-\infty)} = e^{\pm i
x_-(-\infty)}\left(e^{\pm i \pws} -1\right)\,,
\eea
where we take
into account that $x_-(\infty)-x_-(-\infty)=\pws$.

\medskip
Making use of the particular basis described in the previous
section and imposing the boundary condition $x_-(-\infty)=0$,
we identify the exact expression for the central function
$c$ in eqs.(\ref{Beis-alg}): \bea \label{Cc}
c=\frac{1}{2\zeta}(e^{ip_{\rm ws}}-1)\, . \eea
The algebra (\ref{Beis-alg}) with the
central charges of the form (\ref{Cc}) perfectly agrees with the
one considered in \cite{Beisert:2005tm} in the field theory
context.

It is worth noting that there is another natural choice of the
boundary condition for the light-cone coordinate $x_-$: \bea\nonumber
x_-(+\infty)=-x_-(-\infty) = {\pws\ov 2}\,. \eea This is the symmetric
condition which treats both boundaries on the equal footing, and leads
to a purely imaginary central charge \bea \label{Cc2}
c=\frac{i}{\zeta}\sin({\pws\ov 2})\, . \eea It is obvious from
(\ref{Ccg}) that different boundary conditions for $x_-$ lead to
central charges which differ from each other by a phase
multiplication. This freedom in the choice of the central charge
follows from the obvious ${\rm U}(1)$ automorphism of the algebra
(\ref{Beis-alg}): one can multiply all supercharges $Q_a^\a$ by any
phase which in general may depend on all the central charges.

\medskip

Since we already obtained the expected central charges, the
contribution of all the other terms in (\ref{pbsch}) should
vanish. Indeed, the second term in (\ref{uusch}) contributes to the
order $\z$ in the expansion as can be easily seen integrating by parts
and using the relation $x'_- \sim -\zeta px' +\cdots$. Taking into
account the additional contribution to the terms of order $\z$ in
(\ref{pbsch}), we have checked that the total contribution is given by
a $\s$-derivative of a local function of $x$ and $p$, and, therefore,
only contributes to terms of order $\z^2$.

\medskip

We have verified up to the quartic order in fields that the
Poisson bracket of supercharges with $\a_1=-\a_2$ gives the
Hamiltonian and the kinematic generators in complete agreement
with the centrally extended $\su(2|2)$ algebra (\ref{Beis-alg}).

\medskip

The next step is to show that the Hamiltonian commutes with all
dynamical supercharges. As was already mentioned, this can be done
order by order in perturbation theory in powers of the inverse
string tension $\zeta$ and in number of fermionic variables. We
have demonstrated that up to the first non-trivial order $\zeta$
the supercharge ${\rm Q}$ truncated to the terms linear in
fermions indeed commutes with ${\rm H}$. To do that we need to
keep in ${\rm H}$ all quadratic and quartic bosonic terms, and
quadratic and quartic terms which are quadratic in fermions, see
the Appendix for details.

\medskip

The computation we described above was purely classical, and one
may want to know if quantizing the model could lead to some kind
of an anomaly in the symmetry algebra. We have computed the
symmetry algebra in the plane-wave limit where one keeps only
quadratic terms in all the symmetry generators, and shown that all
potentially divergent terms cancel out and no quantum anomaly
arises. As a result, one gets again the same centrally extended
$\su(2|2)$ algebra (\ref{Beis-alg}) with the central charges
$1/\z (e^{\pm i\pws}-1)$ replaced by their low-momentum
approximations $\mp i \int_{-\infty}^\infty {\rm d}\s
\Big(p_Mx'_M-\frac{i}{2}\str(\Sigma_+\chi\chi') \Big)$.

\medskip

Thus, we have shown that in the infinite $P_+$ limit and for physical fields
chosen to rapidly decrease at infinity the corresponding string model
enjoys the symmetry which coincides with two copies of the centrally-extended $\su(2|2)$ algebra (\ref{Beis-alg})
sharing the same Hamiltonian.

%%%%%%%%%%%%%%%%%%%%%%%%%%%%%%%%%%%%%%%%%%%%%%%%%%%%%%%%%%%%%%%%%%%
\section{Concluding remarks}
The main focus of this paper has been on the analysis of the
off-shell string symmetry algebra in the limit of \emph{infinite}
light-cone momentum $P_+$. Relaxing the level-matching condition
brings only one modification in this case: namely, the algebra
$\psu(2|2)\oplus\psu(2|2)$ undergoes extension by a new central
charge proportional to the level-matching condition.

\medskip

The physically more relevant situation, however, corresponds to
the case of a finite light-cone momentum. For $P_+$ finite the
zero mode of the conjugate field $x_-$ has to be taken into
account. Also, since the length of the string is finite,
transverse fields do not have to vanish at the string end points.
So the question arises what is the symmetry algebra in this case?

\medskip

Recall that relaxing the level-matching condition for finite $P_+$
means
\begin{equation}
\nn x_-(r)-x_-(-r)= p_{\text{ws}}\,,   \quad -r \leq \sigma\leq r
\,, \quad  r=\frac{\pi P_+}{\sqrt{\lambda}} \, ,
\end{equation}
which implies that the Poisson bracket of the dynamical
supercharge $\Q$, eq.(\ref{sch}), with  the level-matching
generator is
\begin{equation}
\nn \frac{1}{\zeta}\{p_{\rm ws},{\rm Q}\}  = \int_{-r}^r {\rm
}{\rm d}\sigma~
\partial_\sigma {\rm Q} = \Omega(r)e^{i \alpha x_-(-r)} (e^{i
\alpha p_{\text{ws}}} -1) \neq 0 \, .
 \end{equation}
Hence this Poisson bracket does not vanish, since
$\Omega(r)=\Omega(-r)$ is non-zero in the finite $P_+$ case.
Similarly the Poisson bracket of the same supercharge with the
Hamiltonian will be non-vanishing. Thus, we see that the off-shell
extension of the theory does not allow one to maintain the
$\psu(2,2|4)$ symmetry algebra for a string of finite length.

\medskip

It should be further noted that an off-shell theory is not
uniquely defined. Indeed, one can use the level-matching generator
to modify the Hamiltonian
$$
{\rm H}\to {\rm H}+c_n p_{\rm ws}^n\, ,
$$
where the coefficients $c_n$ might depend on physical fields. On
the states satisfying the condition of level-matching the new
Hamiltonian reduces to the original one. The absence of the
standard $\psu(2,2|4)$ symmetry in an off-shell theory does not a
priory preclude the existence of new hidden symmetries of the
off-shell Hamiltonian. Their discovery would provide a substantial
step in understanding the string dynamics for the physically
relevant situation of the finite light-cone momentum.

%This replacement boils down to substituting in all dynamical
%generators the variable $x_-$ for a new expression $\tilde{x}_-$:
%$$
%x_-\to \tilde{x}_-=x_--\frac{\sigma}{2r}{\rm P}\, .
%$$ The field $\tilde{x}_-$ is periodic and its time derivative
%coincides with that of $x_-$ because $\rm P$ is the integral of
%motion. This property will ensure that the dynamical generators
%constructed in this way will commute with both ${\rm P}$ and ${\rm
%H}$.
%Whether this idea is correct or not requires further
%investigation, in particular understanding the connection with the
%string theory results at finite $P_+$ \cite{Frolov:2004bh},
%\cite{SZZ},\cite{AJK},\cite{magnon}.
\medskip

In  the  hybrid expansion  used  in our  paper,  the  crucial role  in
deriving the non-linear central  charges, was played by the ``vertex''
$e^{i \alpha  x^-}$. The question is  what is the  physical meaning of
this object? To  see this, consider the quantum  theory.  The variable
$x_-(s)$ contains a zero mode \footnote{This is an integration constant
arising  upon  integrating  equation.(\ref{eqxmin}).} $\hat{x}_-$  which   is conjugate to the operator $\hat{P}_+$
$$
[\hat{P}_+,\hat{x}_-]=-i\, .
$$
Thus, if we consider a state $|P_+\rangle$ with a definite value
of $\hat{P}_+|P_+\rangle=P_+|P_+\rangle$ then a state $e^{i\a
\hat{x}_-}|P_+\rangle$ carries a new value of $P_+$:
$$
\hat{P}_+e^{i\a x_-}|P_+\rangle=(\a+P_+)e^{i\a x_-}|P_+\rangle
$$
Since in the light-cone approach $P_+$ is naturally identified
with the length of the string, it is appropriate to call $e^{i\a
x_-}$ the {\it length-changing operator}. The Hilbert space of the
corresponding theory is necessarily a direct sum: ${\mathcal H}=
\sum_{P_+}{\mathcal H}_{P_+}$ of the spaces ${\mathcal H}_{P_+}$
corresponding to an individual eigenvalue of the operator
$\hat{P}_+$.

\medskip
This brief discussion of the light-cone string theory for finite
$P_+$ clearly demonstrates that the latter carries many subtleties
with respect to the infinite $P_+$ limit, which for sure require
further investigation.

%%%%%%%%%%%%%%%%%%%%%%%%%%%%%%%%%%%%%%%%%%%%%%%%%%%%%%%%%%%%%%%%%%%%%%%%

\section*{Acknowledgements}

We wish to thank Niklas Beisert for valuable discussions. The work
of G.~A. was supported in part by the RFBI grant N05-01-00758, by
NWO grant 047017015 and by the INTAS contract 03-51-6346. 
The work
of G.~A. and S.~F.~was supported in part by the EU-RTN network
{\it Constituents, Fundamental Forces and Symmetries of the
Universe} (MRTN-CT-2004-005104). The work of J.~P. is supported by
the Volkswagen Foundation. He also thanks the
Albert-Einstein-Institute for hospitality. 
The work of M.~Z.  was supported in part by the grant {\it Superstring
Theory} (MRTN-CT-2004-512194). M.~Z. would like to
thank the Spinoza Insitute for the hospitality during the last
phase of this project.

%%%%%%%%%%%%%%%%%%%%%%%%%%%%%%%%%%%%%%%%%%%%%%%
\section{Appendix}
%%%%%%%%%%%%%%%%%%%%%%%%%%%%%%%%%%%%%%%%%%%%%%%

\subsection{The gauge-fixed Hamiltonian}
\label{appA}
The Hamiltonian for physical excitations arising in the light-cone
gauge was found in \cite{LCpaper}. The gauge choice made in
\cite{LCpaper} is however not exactly the same as eq.(\ref{gauge})
adopted here. Also the theory in \cite{LCpaper} is defined on the
standard interval for $\sigma$: $-\pi \leq\sigma\leq \pi$. In
order to make a connection with the results by \cite{LCpaper} we
choose the variable $p_+$ there to be equal to $p_+=2P_+$, where
$P_+$ is identified with the total momentum in \eqn{Pplusdef}. To
justify our choice we note that with $p_+=2P_+$ the gauge-fixed
action of \cite{LCpaper} can be schematically represented in the
form \bea \label{FPSaction} {\rm S}=P_+
\int_{-\pi}^{\pi}\frac{{\rm d}\sigma {\rm d}\tau
}{2\pi}\Big(p_M{\dot x}_M+\chi^{\dagger}\dot{\chi}-{\cal H}\Big)\,
, \eea where ${\cal H}$ is the $P_+$-independent Hamiltonian
density, $(x_M,p_M)$ with $M=1,\ldots, 8$ are transverse
coordinates and their conjugate momenta, and $\chi$ encodes the
fermionic variables. Since we are interested in the limit of
infinite $P_+$, it is appropriate to make a rescaling $\sigma\to
\frac{\sqrt{\lambda}}{P_+}\sigma$. Upon this rescaling the action
(\ref{FPSaction}) turns precisely into eq.(\ref{Sgf}) of the
present paper with $r=\frac{\pi}{\sqrt{\lambda}}P_+$. As was
discussed in section 2, we further supplement this rescaling of
$\sigma$ with rescalings of physical variables \bea \label{res}
(x_M,p_M,\chi)\to \sqrt{\zeta}(x_M,p_M,\chi)\, . \eea This is
necessary  in order to ensure to have canonical Poisson brackets
for physical fields.

\medskip

Upon these redefinitions the Hamiltonian found in \cite{LCpaper} turns into the one given by eq.(\ref{structureH})
Explicitly, the quadratic piece of the Hamiltonian density in eq.(\ref{structureH}) has the form
\bea \label{H2} {\cal
H}_2=\frac{1}{2}p_M^2+\frac{1}{2}x_M^2+\frac{1}{2}x'^2_M+\frac{1}{2}
\str(\Sigma_+\chi\tilde{K}\chi'^tK )+\frac{1}{2}\str~\chi^2\,
,~~~~~~~~ \eea while the quartic one is \cite{LCpaper} \bea
\nonumber && ~~~~~~~~~{\cal H}_4=\frac{1}{4}\Big[p_y^2 z^2-p_z^2
y^2 +(y'^2z^2-z'^2y^2)+2(z'^2z^2-y'^2y^2)+\\  \nonumber
&&~~~~~~~~~~~~~~~~~~~+\str\Big(
(z^2-y^2)\chi'\chi'+\frac{1}{2}[\Sigma'(x),\Sigma(x)](\chi\chi'-\chi'\chi)-2\Sigma(x)\chi'\Sigma(x)\chi'\Big)\\
&&~~~~~~~~~~~~~~~~~~~+\frac{i}{4}
\str\Big([\Sigma(x),\Sigma(p)]'(\tilde{K}\chi^tK\chi-\chi\tilde{K}\chi^t
K)\Big)~\Big]+{\cal O}(\chi^4)\, , \label{H4}\eea where by ${\cal
O}(\chi^4)$ we encode all the terms which are quartic in fermions
(stated in \cite{LCpaper}). The transverse bosonic fields we have
denoted as $x_M=(z_a,y_s)$ with $z_a$ ($a=1,2,3,4$) accounting for
the transverse ${\rm AdS}_5$ and $y_s$ ($s=1,2,3,4$) for the $S^5$
degrees of freedom. Prime denotes $\partial_\sigma$ and in the
fermionic sector we have introduced the following notation \bea
\nonumber K = \left(
\begin{array}{cc}
  K_4 & 0  \\
  0 & K_4
\end{array} \right)\, ,\qquad \widetilde{K} = \left(
\begin{array}{cc}
  K_4 & 0  \\
  0 & -K_4
\end{array} \right)\, .
\eea
with the matrix $K_4$ satisfying $K_4^2=-{\mathbb I}$ given by
\bea
\nonumber
K_4={\scriptsize\left(
\begin{array}{cccc}
  0 & 1 & 0 & 0 \\
  -1 & 0 & 0 & 0 \\
   0 & 0 & 0 & 1 \\
   0 & 0 & -1 & 0
\end{array} \right)\, .}
\eea
We also use the notation $\Sigma(x)=\Sigma_M x_M$ and $\Sigma(p)=\Sigma_M p_M$. The
$8\times 8$ matrices $\Sigma_M$ have the following structure
\bea\la{SM}
\S_M= \left\{\left(
\begin{array}{cc}
 \gamma_a& 0  \\
  0 & 0
\end{array} \right)\,, \left(
\begin{array}{cc}
 0& 0  \\
  0 & i  \gamma_s
\end{array} \right)\right\}\,
\eea and are written in terms of the four Dirac matrices
$\gamma_i$. We work with the basis defined in appendix A of
\cite{LCpaper}. For the definition of the matrices $\Sigma_{\pm}$
see eq.(\ref{Sigmapm}).

\medskip

The fermions enter in the above through
 the $\kappa$-gauge fixed $8\times 8$ matrix
$\chi$ (compare (A.6) and (A.9) of \cite{LCpaper})
\begin{equation}
\chi = \left ( \begin{matrix}
0 & {\cal P}_+\,\eta + {\cal P}_- \,\theta^\dagger \cr
-  \cal P_-\, \eta^\dagger + {\cal P}_+\, \theta & 0
\end{matrix} \right )\, ,
\qquad {\cal P}_+= \left (
\begin{matrix} {\mathbb I}_2 & 0 \cr 0 & 0
\end{matrix}\right ) \, ,
\qquad
 {\cal P}_-= \left (
\begin{matrix} 0&0 \cr 0 & {\mathbb I}_2 
\end{matrix}\right )
\end{equation}
where \be \label{ferms1} \eta=\sum_{i=1}^4 \tilde\eta_i\,
\Gamma_i\, , \qquad \theta=\sum_{i=1}^4\tilde\theta_i\, \Gamma_i
\, , \ee with the Dirac matrices $\Gamma_i$ in the complex basis
defined in \cite{LCpaper}, explicitly \bea \Gamma_1 &=
\sfrac{1}{2}(\gamma_2-i\gamma_1) =  \left (
\begin{matrix} 0&0&0& i\cr 0&0&0&0\cr 0&-i&0&0\cr 0&0&0&0
\end{matrix}\right )\, , \qquad
\nn \Gamma_2 &= \sfrac{1}{2}(\gamma_4-i\gamma_3) = 
\left (
\begin{matrix} 0&0&-i& 0\cr 0&0&0&0\cr 0&0&0&0\cr 0&-i&0&0
\end{matrix}\right )\, \eea
and $ \Gamma_4=(\Gamma_1)^\dagger$, $\Gamma_3=(\Gamma_2)^\dagger$.
Moreover we define the standard double index Dirac matrices by
$\Gamma_{ab} := \sfrac{1}{2}\, [\Gamma_a,\Gamma_b] $. We also define two-dimensional projectors
\bea
P^+_2= \left(
\begin{array}{cc}
  1 & 0 \\
  0 & 0 
\end{array} \right)\, ;\quad P^-_2 = \left(
\begin{array}{cc}
  0 & 0 \\
  0 & 1 
\end{array} \right)\, .
\eea

\subsection{Symmetry generators}
To describe the symmetry generators $\Q$ and the gauge-fixed Hamiltonian we have to
introduce a proper parametrization of the coset space $
\frac{\rm PSU(2,2|4)}{{\rm SO(4,1)}\times {\rm SO(5)}}\, $. Following \cite{LCpaper} we chose the coset
representative in the form
$$
g(\chi,x,t,\phi)=\Lambda(t,\phi)g(\chi)g(x)\, .
$$
Here $x_M =(z_a,y_i)$ is a short-hand notation for the transverse
bosonic fields and $\chi$ denotes the 16 physical fermions which
are left upon fixing the $\kappa$-symmetry. The matrix
$\Lambda(t,\phi)$ was defined in \eqn{lambdadef}. The element
$g(x)$ is the $8\times 8$ matrix which has the following structure
in terms of $4\times 4$ blocks related to the AdS and to the
sphere parts respectively \bea \nonumber
g(x)=\left(\begin{array}{cc}
{1\over \sqrt{1- \zeta {z^2 \over 4 }}} \Big(1 + {\sqrt{\zeta} \over 2 } z_a \gamma_a\Big) ~&~ 0  \\
0 ~&~ {1\over \sqrt{1 + \zeta {y^2 \over 4 }}} \Big(1 + {i \sqrt{\zeta} \over 2} y_i \gamma_i\Big)  \\
\end{array}\right) \, .
\eea Finally the fermionic coset element reads \cite{LCpaper}
\be\nn g(\chi) = \sqrt{\zeta}\,\chi +\sqrt{1+\zeta\, \chi^2} \, .
\ee The $8\times 8$ supermatrix $\Q$ of \eqn{charges} is then
defined by \cite{LCpaper} \bea\la{Qconcrete} {\rm Q}
=\int_{-r}^{r} {\rm d}\sigma~ \Lambda U \Lambda^{-1}\, , ~~~~~
\eea where \bea \label{U} U=g(\chi) g(x)\left(\bp +{i\ov
2}g(x)\widetilde{K} F_\s^t K g(x)^{-1} \right)
g(x)^{-1}g(\chi)^{-1}\, . \eea Here $K$ and $\tilde K$ have been
defined above. We also have \be\nn F_\s =
\sqrt{\zeta}\,(\sqrt{1+\zeta \chi^2}\pa_\s\chi -
\chi\pa_\s\sqrt{1+\zeta\,\chi^2}\,) \, \ee and $\bp$ is defined by
\bea
%\label{bpexp}
\nn \bp  = {i\ov 4}\bp_+ \Sigma_+ +
 {i\over 4}\bp_- \Sigma_- + {1\over 2}\bp_M \Sigma_M \, ,
\eea where
\begin{align}\nn
\bp_+ &= \frac{1}{G_+} (2 + G_- \bp_-)\, ,\qquad
\bp_- = - { G_+(\bp_M^2 + \, {\cal A}^2) \ov  (1 + \sqrt{ 1 - G_+G_-(\bp_M^2 +\, {\cal A}^2)})}\, , \nn\\
\nn {\bp_a} &={\sqrt{\zeta}}\,  p_a (1 - \zeta {z^2\ov 4})\,
,\qquad {\bp_s} = {\sqrt{\zeta}}\,  p_s (1 + \zeta {y^2\ov 4})\, .
\end{align}
Finally ${\cal A}^2$ and $G_\pm$ are given by
\begin{equation}\nn
{\cal A}^2 = - x_-'^2\,G_+G_-  +{ \zeta z_a'^2  \ov  \left( 1-
\zeta {z^2\ov 4} \right)^2} + {\cal O}(\chi^2) \, , \qquad G_\pm
={1\ov 2}\left( {1+ \zeta {z^2\ov 4}\ov 1- \zeta {z^2\ov 4} } \pm
{1- \zeta {y^2\ov 4}\ov 1+ \zeta {y^2\ov 4}}\right) \, .
\end{equation}

\subsection{Covariant notation}

As was done in \cite{LCpaper} we shall make use of complex fields.
However, here we will denote them in a covariant notation with
upper and lower indices reflecting their charges under the four
transverse ${\rm U}(1)$ subgroups involved. The bosonic fields we
denote by
\begin{align}
Z_1&= z_2+iz_1 \, ; \quad Z_2=z_4+iz_3 \, ; \quad Z^2=(Z_2)^\dagger \, ; \quad
Z^1=(Z_1)^\dagger\, ; \nn\\
Y_1&= y_2+iy_1 \, ; \quad Y_2=y_4+iy_3 \, ; \quad Y^2=(Y_2)^\dagger\, ; \quad
Y^1=(Y_1)^\dagger\, ; \nn \\
P^Z_1 &= \sfrac{1}{2}(p^z_2+ip^z_1)\, ; \quad  P^Z_2=\sfrac{1}{2}(p^z_4+ip^z_3) \, ; \quad
(P^Z)^1= (P^z_1)^\dagger \, ; \quad (P^Z)^2= (P^Z_2)^\dagger \, ;\nn\\
P^Y_1 &= \sfrac{1}{2}(p^y_2+ip^y_1)\, ; \quad
P^Y_2=\sfrac{1}{2}(p^y_4+ip^y_3) \, ; \quad (P^Y)^1=
(P^Y_1)^\dagger \, ; \quad (P^Y)^2= (P^Y_2)^\dagger \, . \nn
\end{align}
with the quantum commutation relations ($\alpha,\beta=1,2$ and $a,b=1,2$)
\begin{equation}
\begin{aligned}
\label{comrels1}
 &[P^Z_\alpha(\sigma) , Z^\beta(\sigma')]
 =-i\,\delta^\beta_\alpha\,\delta(\sigma-\sigma')~~
&[(P^Z)^\alpha(\sigma) , Z_\beta(\sigma')] =-i\,\delta_\beta^\alpha\,\delta(\sigma-\sigma') \\
 & [P^Y_a(\sigma) , Y^b(\sigma')] =-i\,\delta^b_a \,
\delta(\sigma-\sigma') ~~ &[(P^Y)^a(\sigma) , Y_b(\sigma')]
=-i\,\delta_b^a\, \delta(\sigma-\sigma')\, ,
\end{aligned}
\end{equation} analogue expressions apply at the classical level
for the Poisson brackets.

We also introduce upper and lower indices for the fermionic fields
defined in \eqn{ferms1} by denoting \begin{equation}
\begin{aligned}
\tilde\theta_1&=\theta_1 ~~&~~  \tilde\theta_2&=\theta_2 ~~&~~
\tilde\theta_3&=\theta^2
~~&~~ \tilde\theta_4&=\theta^1 \\
\tilde\theta^\dagger_1&=\theta^{\dagger\, 1} &
\tilde\theta^\dagger_2&=\theta^{\dagger\, 2}
&\tilde\theta^\dagger_3&=\theta^\dagger_2
 &\tilde\theta^\dagger_4&=\theta^\dagger_1 \\
\tilde\eta_1&=\eta_1 & \tilde\eta_2&=\eta_2  &
\tilde\eta_3&=\eta^2
& \tilde\eta_4&=\eta^1 \\
\tilde\eta^\dagger_1&=\eta^{\dagger\, 1} &
\tilde\eta^\dagger_2&=\eta^{\dagger\, 2}  &
\tilde\eta^\dagger_3&=\eta^\dagger_2  &
\tilde\eta^\dagger_4&=\eta^\dagger_1 \, ,
\end{aligned}
\end{equation}
leading to the covariant anti-commutation relations
\begin{equation}
\begin{aligned}
\label{comrels2} &\{\theta_\alpha(\sigma),\theta^{\dagger\,
\beta}(\sigma')\} =\delta^\beta_\alpha\, \delta(\sigma-\sigma')
~~&~~
&\{\theta^\alpha(\sigma),\theta^\dagger_\beta(\sigma')\} =\delta_\beta^\alpha\,\delta(\sigma-\sigma')\\
&\{\eta_a(\sigma),\eta^{\dagger\, b}(\sigma')\} =\delta^b_a\,
\delta(\sigma-\sigma')   &
&\{\eta^a(\sigma),\eta^\dagger_b(\sigma')\} =\delta_b^a\,
\delta(\sigma-\sigma') \, .
\end{aligned}
\end{equation}

It is useful to note the charges carried by the fields of the four
${\rm U}(1)$ subgroups involved. For this consider the
combinations $S_\pm = S_1 \pm S_2$ and $J_\pm = J_1\pm J_2$. Then
the $\su(2|2)_R$ right (blue) generators carry $J_-$ and $S_-$
charges whereas the $\su(2|2)_L$ left (red) generators are charged
under $S_-$ and $J_-$. The following tables exemplify this:
$$
\begin{array}{|c|r|r|r|r|}
\hline
&S_+&S_-&J_+&J_-\\
\hline
~Z_1,\, (P^Z)_1,\, \bar{Z}_1,\, (\bar{P}^Z)_1,\,~&1&~1&~0&~0\\
\hline
~Z_2,\, (P^Z)_2,\, \bar{Z}^2,\, (\bar{P}^Z)^2,\,~&~1&-1&~0&~0\\
\hline
~Z^2,\, (P^Z)^2,\, \bar{Z}_2,\, (\bar{P}^Z)_2,\,~&-1&~1&~0&~0\\
\hline
~Z^1,\, (P^Z)^1,\, \bar{Z}^1,\, (\bar{P}^Z)^1,\,~&-1&-1&~0&~0\\
\hline
\end{array}\,,\quad
\begin{array}{|c|r|r|r|r|}
\hline
&S_+&S_-&J_+&J_-\\
\hline
~Y_1,\, (P^Y)_1,\, \bar{Y}_1,\, (\bar{P}^Y)_1,\,&~0&~0~&1&~1\\
\hline
~Y_2,\, (P^Y)_2,\, \bar{Y}^2,\, (\bar{P}^Y)^2,\,&~0&~0~&~1&-1\\
\hline
~Y^2,\, (P^Y)^2,\, \bar{Y}_2,\, (\bar{P}^Y)_2,\,&~0&~0~&-1&~1\\
\hline
~Y^1,\, (P^Y)^1,\, \bar{Y}^1,\, (\bar{P}^Y)^1,\,&~0&~0~&-1&-1\\
\hline
\end{array}
$$

$$
\begin{array}{|c|r|r|r|r|}
\hline
&S_+&S_-&J_+&J_-\\
\hline
~\theta_1,\, \theta^\dagger_1,\, \bar{\theta}_1,\, \bar{\theta}^\dagger_1,\,~&0&1&1&~0\\
\hline
~\theta_2,\, \theta^\dagger_2,\, \bar{\theta}^2,\, \bar{\theta}^{\dagger\, 2},\,~&0&-1&1&~0\\
\hline
~\theta^2,\, \theta^{\dagger\, 2},\, \bar{\theta}_2,\, \bar{\theta}^\dagger_2,\,~&0&1&-1&~0\\
\hline
~\theta^1,\, \theta^{\dagger\, 1},\, \bar{\theta}^1,\, \bar{\theta}^{\dagger\, 1},\,~&0&-1&-1&~0\\
\hline
\end{array}\,,\quad
\begin{array}{|c|r|r|r|r|}
\hline
&S_+&S_-&J_+&J_-\\
\hline
~\eta_1,\, \eta^\dagger_1,\, \bar{\eta}_1,\, \bar{\eta}^{\dagger}_1,\,~&1&0&0&1\\
\hline
~\eta_2,\, \eta^\dagger_2,\, \bar{\eta}^2,\, \bar{\eta}^{\dagger\, 2},\,~&1&0&0&-1\\
\hline
~\eta^2,\, \eta^{\dagger\, 2},\, \bar{\eta}_2,\, \bar{\eta}^\dagger_2,\,~&-1&0&0&1\\
\hline
~\eta^1,\, \eta^{\dagger\, 1},\, \bar{\eta}^1,\, \bar{\eta}^{\dagger\, 1},\,~&-1&0&0&-1\\
\hline
\end{array}
$$
Hence a lower (upper) index on
$Z,Y,P^Z,P^Y,\theta,\eta,\theta^\dagger$ and $\eta^\dagger$
denotes a charge of 1 (-1) with respect to $(S_++J_+)$. In the
above tables we have also introduced barred coordinates defined
with a flipped ``2'' index as \be \la{barred} \bar A_1 =A_1 \, ,
\quad \bar A^1 =A^1\, , ~~~ \bar A_2 =A^2\, , ~~~ \bar A^2 =A_2 \,
,~~~ \mbox{with}~~~ A\in
\{Z,Y,P^Z,P^Y,\theta,\eta,\theta^\dagger,\eta^\dagger\} \, , \nn
\ee which are the natural objects for the $\su(2|2)_L$ left (red)
generators as we shall see shortly. For the barred coordinates the
index position now denotes the charge with respect to $(S_-+J_-)$.
Clearly the commutation relations keep their canonical form, cf.
\eqn{comrels1} and \eqn{comrels2}, in the barred coordinates.

\subsection{The explicit form of the $\su(2|2)_R$ generators}

Using the basis of fermionic (dynamical) generators of the right
(blue) $\su(2|2)_R$ algebra given in \eqn{BcB} and \eqn{Bcc} along
with the concrete expression for $\Q$ in \eqn{Qconcrete} one finds
the leading quadratic order expressions for
the supercharges 
\begin{align}
Q^\alpha{}_a &= -\frac{1}{2}\,\int {\rm d}\sigma\,
e^{-\frac{i}{2}\,x_-}\Bigl [ i\, \theta^\alpha\, (2P^Y+i Y)_a
+(2P^Z-i Z)^\alpha\, \eta^\dagger_a -\theta^{\dagger\,\alpha}\,
Y'_a - i Z'^\alpha\,\eta_a  \nonumber \\
&\quad + \epsilon^{\alpha\beta}\epsilon_{ab}\, \Bigl (i
\theta_\beta\, (2P^Y + i Y)^b +
(2P^Z-i Z)_\beta \,\eta^{\dagger\, b} - \theta^\dagger_\beta\, Y'^b-i\, Z'_\beta\,\eta^b\Bigr ) \, \Bigr ]\\ \nn\\
\bar Q_\alpha{}^a &= \frac{1}{2}\,\int {\rm d}\sigma\,
e^{\frac{i}{2}\, x_-}\, \Bigl [ i\theta^\dagger_\alpha\,
(2P^Y-iY)^a -(2P^Z+iZ)_\alpha\, \eta^a +\theta_{\alpha}\,
Y'^a - i Z'_\alpha\,\eta^{\dagger a}   \nonumber \\
&\quad +\epsilon_{\alpha\beta}\epsilon^{ab}\, \Bigl (i
\theta^{\dagger\,\beta}\, (2P^Y-iY)_b - (2P^Z+iZ)^\beta \,\eta_{b} +
\theta^\beta\, Y'_b-i\, Z'^\beta\,\eta^\dagger_b\Bigr ) \, \Bigr ] =(Q^\alpha{}_a)^\dagger\, .
\end{align}
Moreover the $\su(2)$ generators $R^\alpha{}_\beta$ and
$L^a{}_b$ can be computed using the basis of  (\ref{BL}) and
(\ref{BR}). They read 
\begin{align}
R^\alpha{}_\beta &= \int {\rm d}\sigma\, \Bigl (i [(P^Z)^\alpha\,
Z_\beta -(P^Z)_\beta\, Z^\alpha] + \frac{i}{2}\,
\delta^\alpha_\beta
\, [ (P^Z)_\gamma\, Z^\gamma- (P^Z)^\gamma\, Z_\gamma ] \nn\\
&~~~+ \theta^\dagger_\beta\, \theta^\alpha - \theta^{\dagger\, \alpha}\, \theta_\beta + \frac{1}{2}\,
\delta^\alpha_\beta\, [ \theta^{\dagger\,\gamma}\, \theta_\gamma - \theta^\dagger_\gamma\, \theta^\gamma ]\,
\Bigr ) \la{Rr}\\
L^a{}_b &= \int{\rm d} \sigma\,  \Bigl (\, i[(P^Y)^a\,  Y_b
-(P^Y)_b\, Y^a] + \frac{i}{2}\, \delta^a_b
\, [ (P^Y)_c\, Y^c- (P^Y)^c\, Y_c ] \nn\\
&~~~+ \eta^\dagger_b\, \eta^a - \eta^{\dagger\, a}\, \eta_b + \frac{1}{2}\,
\delta^a_b\, [ \eta^{\dagger\,c}\, \eta_c - \eta^\dagger_c\, \eta^c ] \, \Bigr )\la{Lr} \, .
\end{align} 
Using the above expressions for the supersymmetry generators, it
is straightforward to compute their quantum anti-commutators. One
indeed finds \be \la{QrBQr} \{\, Q^\alpha{}_a , \bar Q_\beta{}^b
\, \} = \delta^b_a\, R^\alpha{}_\beta + \delta^\alpha_\beta\,
L^b{}_a + \frac{1}{2}\delta^b_a\, \delta^\alpha_\beta\, {\rm H} \ee 
with the
Hamiltonian
\begin{align}
\la{Hquantum} {\rm H}&= 2\int {\rm d}\sigma\, \Bigl [(P^Z)^\gamma\,
(P^Z)_\gamma +(P^Y)^c\, (P^Y)_c + \frac{1}{4}\,
(Z^\gamma Z_\gamma +Z'^\gamma Z'_\gamma + Y^c Y_c + Y'^c Y'_c) \nn\\
&~+\frac{1}{2} ( \theta^{\dagger\,\gamma}\theta_\gamma +
\theta^{\dagger}_\gamma\theta^\gamma
 +\eta^{\dagger\, c} \,\eta_c+\eta^{\dagger}_c \,\eta^c)
+\frac{1}{2} (\theta^{\dagger\prime}_\gamma\theta^{\dagger\,\gamma} -\theta^{\prime\,\gamma}\theta_\gamma
 +\eta^{\dagger\prime}_c \,\eta^{\dagger\, c}-\eta^{\prime\, c} \,\eta_c) -2\,\delta(0)\Bigr ]\, .
\end{align}
The normal ordering contribution $-2\,\delta(0)$ of the fermions
will cancel against the ground state energy of the bosons by
supersymmetry upon introduction of creation and annihilation
operators.

The  $\su(2)$ generators $R^\alpha{}_\beta$ and $L^a{}_b$
(\ref{Lr}) and (\ref{Rr}) can be shown to obey the commutation
relations \be \la{RRLL} [R^\alpha{}_\beta , R^\gamma{}_\delta ] =
\delta^\gamma_\beta\,  R^\alpha{}_\delta -  \delta^\alpha_\delta\,
R^\gamma{}_\beta \, ,\qquad [L^a{}_b , L^c{}_d ] = \delta^c_b\,
L^a{}_d -  \delta^a_d\, L^c{}_b \, . \ee Next we turn to the
quantum anticommutator $\{Q^\alpha{}_a, Q^\beta{}_b\}$ which
evaluates to 
\begin{equation}
%\la{QrQr}
\nn \{Q^\alpha{}_a, Q^\beta{}_b\} = \frac{i}{2\zeta} \,
\epsilon^{\alpha\beta}\, \epsilon_{ab}\,\int {\rm d}\sigma\,
e^{-i\, x_-}\, x'_- + \int {\rm d}\sigma_1\, {\rm d}\sigma_2\,
\delta(\sigma_1-\sigma_2)\,
(\partial_{\sigma_1}+\partial_{\sigma_2})\,
\delta(\sigma_1-\sigma_2)
\end{equation}
We see that the potential last quantum anomaly cancels and we
recover the central charge announced in \eqn{Beis-alg} also at the
quantum level. The analogous computation for  $\{\bar
Q^\alpha{}_a, \bar Q^\beta{}_b\}$ follows from conjugation.

\medskip

Finally let us stress that in the above computations we have
freely performed partial integrations by dropping contributions
arising from the vertex operators $e^{\pm ix_-}$, as these would
take us beyond the leading quadratic field
approximation. These terms where dealt with, however, at the
classical level up to order ${\cal O}(\zeta)$ as discussed in
section four.

\subsection{The $\su(2|2)_R$ supercharge at quartic field order}

Here we spell out the contribution to the right (blue)
supercharges at quartic field order (${\cal O}(\zeta^2)$)
explicitly, restricting to the terms linear in fermions whose
Poisson brackets yield the quartic bosonic Hamiltonian
\begin{align}
Q^a{}_\alpha \Bigr |_{fbbb} &= \int {\rm d}\sigma e^{-i\, x_-/2}\,
\Bigl \{\nn\\& \phantom{+ } (\theta^\alpha\, Y_a
+\epsilon^{\alpha\beta}\,\epsilon_{ab}\, \theta_\beta\, Y^b)\,
[-\sfrac{i}{4}\,(P^Y)\circ Y
-\sfrac{1}{2}\, {\cal H}_{\rm bos}] \nn\\
& + \Bigl (\theta^\alpha\, (2P^Y-iY)_a +\epsilon^{\alpha\beta}\,\epsilon_{ab}\, \theta_\beta\,
 (2P^Y-iY)^b\, \Bigr ) \, [\sfrac{i}{4}\, Y\circ Y] \nn\\
&+(\theta^{\dagger\, \alpha}\, Y_a+ \epsilon^{\alpha\beta}\,\epsilon_{ab}\, \theta^\dagger_\beta\,
Y^b\, )\, [\sfrac{i}{4}\, (P^Y\circ Y'+P^Z\circ Z')+\sfrac{1}{4}\, Z\circ Z'
-\sfrac{1}{8}\, Y\circ Y'] \nn\\
&+(\theta^{\dagger\, \alpha}\, Y^\prime_a+ \epsilon^{\alpha\beta}\,\epsilon_{ab}\,
\theta^\dagger_\beta\, Y^{\prime\, b}\, ) \, [\sfrac{1}{2}\, Z\circ Z -\sfrac{1}{4}\,
Y\circ Y] \nn\\
& + (\eta^\dagger_a\, Z^\alpha +\epsilon^{\alpha\beta}\,\epsilon_{ab}\, \eta^{\dagger\, b}\,
Z_\beta)\, [-\sfrac{1}{4}\,(P^Z)\circ Z
+\sfrac{i}{2}\, {\cal H}_{\rm bos}] \nn\\
& + \Bigl (\eta^\dagger_a\, (2P^Z+iZ)^\alpha +\epsilon^{\alpha\beta}\,\epsilon_{ab}\,
\eta^{\dagger\, b}\, (2P^Z+iZ)_\beta\, \Bigr ) \, [\sfrac{1}{4}\, Z\circ Z] \nn\\
&+(\eta_a\, Z^\alpha+ \epsilon^{\alpha\beta}\,\epsilon_{ab}\, \eta^b\,
Z_\beta\, )\, [\sfrac{1}{4}\, (P^Y\circ Y'+P^Z\circ Z')-\sfrac{i}{4}\, Y\circ Y'
+\sfrac{i}{8}\, Z\circ Z'] \nn\\
&+(\eta_a\, Z^{\prime\,\alpha}+ \epsilon^{\alpha\beta}\,\epsilon_{ab}\,
\eta^b\, Z^{\prime}_\beta\, ) \, [-\sfrac{i}{2}\, Y\circ Y +\sfrac{i}{4}\,
Z\circ Z] \,
\Bigr \}\, ,
\end{align}
where we have used the notation $(P^Z)\circ Z := (P^Z)_\gamma\,
Z^\gamma + (P^Z)^\gamma\, Z_\gamma$ and $(P^Y)\circ Y := (P^Y)_c\,
Y^c + (P^Y)^c\, Y_c$, etc. Also ${\rm H}_{\rm bos}$ denotes the
bosonic part of the free (quadratic) Hamiltonian \eqn{Hquantum}.
Similar expression follow for the left (red) supercharges.

\subsection{The explicit form of the $\su(2|2)_L$ generators}

We denote all the generators appearing in the left (red) $\su(2|2)_L$ algebra by lower
case letters (with the exception of the common central charges).
For the left (red) supercharges we take as a basis
\begin{align}
q^1{}_{1} &= \frac{1}{2}\, \str\, {\rm Q}\, \sigma^+\otimes (\Gamma_{14}+\Gamma_{23} + {\cal P}_+) \nn\\
q^2{}_{2} &= -\frac{1}{2}\,
\str\, {\rm Q}\, \sigma^+\otimes (\Gamma_{14}+\Gamma_{23} - {\cal P}_+) \nn\\
q^1{}_{2} &= \str \, {\rm Q}\, \sigma^+\otimes \Gamma_{12}\nn\\
q^2{}_{1} &= \str \, {\rm Q}\, \sigma^+\otimes \Gamma_{34}\, .
\end{align}
One then finds at quadratic field order{\footnotesize
\begin{align}
 q^1{}_{1} &= \frac{1}{2} \,\int {\rm d}\sigma\, \, e^{ix_-/2}\, \Bigl [
(2P^Z+iZ)^\gamma\, \theta_\gamma + i \, Z'^\gamma\, \theta^\dagger_\gamma +
 i(2P^Y-iY)_c\, \eta^{\dagger\, c} + Y'_c\,\eta^c
\Bigr ] \nn\\
q^2{}_{2} &= \frac{1}{2}\, \,\int {\rm d}\sigma\, e^{ix_-/2}\,
\Bigl [ (2P^Z+iZ)_\gamma\, \theta^\gamma + i \,
Z'_\gamma\,\theta^{\dagger\, \gamma} + i(2P^Y-iY)^c\,
\eta^{\dagger}_c + Y'^c\,\eta_c
\Bigr ] \nn\\
q^1{}_{2} &= \frac{1}{2} \,\int {\rm d}\sigma\, e^{ix_-/2}\,\Bigl
[  \epsilon_{\alpha\beta} \Bigl ((2P^Z+iZ)^{\alpha}\,
\theta^{\beta} + i \, Z'^{\alpha}\, \theta^{\dagger\, \beta}\,
\Bigr ) - \epsilon_{ab}\,\Bigl (
 i(2P^Y-iY)^{a}\, \eta^{\dagger\, b} + Y'^{a}\,\eta^{b} \,\Bigl )\,\Bigr ]
\nn\\
q^2{}_{1} &= -\frac{1}{2} \,\int {\rm d}\sigma\, e^{ix_-/2}\,
\Bigl [\,\epsilon^{\alpha\beta}\Bigl ( (2P^Z+iZ)_{\alpha}\,
\theta_{\beta} + i \, Z'_{\alpha}\, \theta^{\dagger}_{\beta}\,
\Bigr ) - \epsilon^{ab}\, \Bigl (
 i(2P^Y-iY)_{a}\, \eta^{\dagger}_{b} + Y'_{a}\,\eta_{b} \,
\Bigr )\, \Bigr ]
\label{Qls}
\end{align}}%
and their complex conjugated partners $\bar{q}_A{}^{B}$. These
generators can be shown to anti-commute with the right (blue)
supercharges $Q^\alpha{}_a$ and $\bar{Q}_\alpha{}^a$, their
commutation with the right (blue) $\su(2)$ generators
$R^\alpha{}_\beta$ and $L^a{}_b$ is manifest due to  the (right)
covariant notation.

\medskip

Translating these charges into the barred coordinates with the
flipped ``2'' index of \eqn{barred} enables one to write the left
(red) supercharges covariantly 
\begin{align}
q^A{}_B = \frac{1}{2}&\int d\sigma \, e^{ix_-/2}\,  \Bigl  [
(2\bar{p}^Z+i\, \bar{Z})^A\, \bar{\theta}_B + i\, \bar{Z}^{\prime\, A}\, \bar{\theta}^\dagger_B
+ i(2\bar{p}^Y-i\, \bar{Y})_B\, \bar{\eta}^{\dagger\, A} +  \bar{Y}^{\prime}_{B}\, \bar{\eta}^A \nn\\
&+ \epsilon^{AC}\, \epsilon_{BD}\, \Bigl (
\, (2\bar{p}^Z+i\, \bar{Z})_C\, \bar{\theta}^D + i\, \bar{Z}^{\prime}_{C}\, \bar{\theta}^{\dagger\, D}
+
 i(2\bar{p}^Y-i\, \bar{Z})^D\, \bar{\eta}^\dagger_C + \bar{Y}^{\prime\, D}\, \bar{\eta}_{C}
\Bigr )\, \Bigr ]
\end{align}
and the complex conjugate expression $\bar{q}_A{}^B$. Here we
note the conjugation properties $(\bar{Z}_A)^\dagger = \bar{Z}^A$,
$(\bar{\theta}_A)^\dagger=\bar{\theta}^{\dagger\, A}$,
$(\bar{p}^Z_A)^\dagger = (\bar{p}^Z)^A$, etc. One then computes
the anti-commutator
\begin{equation}
\la{Lalg} \{ q^A{}_{B} , \bar{q}_C{}^{D} \} = -\delta^A_C\,
l^{D}_{B} - \delta^{D}_{B}\, r^A_C + \frac{1}{2}\delta^A_C\, \delta^{D}_{B}\,
{\rm H}
\end{equation}
with the same quadratic Hamiltonian $\rm H$ appearing in the
$\su(2|2)_L$ algebra. The bosonic $\su(2)$ generators appearing on
the right hand side are given by
\begin{align}
r^A{}_B &= i\,\Bigl [\,  (\bar{p}^Z)^A\, \bar{Z}_B - (\bar{p}^Z)_B\, \bar{Z}^A\,\Bigr ]
+\frac{i}{2}\, \delta^A_B\, \Bigl [ (\bar{p}^Z)_C\, \bar{Z}^C -(\bar{p}^Z)^C\, \bar{Z}_C
\,\Bigr ] \nn \\
& \quad
- \Bigl [\,  (\bar{\eta}^\dagger)^A\, \bar{\eta}_B - (\bar{\eta}^\dagger)_B\, \bar{\eta}^A\,
\Bigr ]
+\frac{1}{2}\, \delta^A_B\, \Bigl [ (\bar{\eta}^\dagger)^C\, \bar{\eta}_C
-(\bar{\eta}^\dagger)_C\, \bar{\eta}^C\, \Bigr ] \, ,\\ \nn \\
l^A{}_B &= i\,\Bigl [\,  (\bar{p}^Y)^A\, \bar{Y}_B - (\bar{p}^Y)_B\, \bar{Y}^A\,\Bigr ]
+\frac{i}{2}\, \delta^A_B\, \Bigl [ (\bar{p}^Y)_C\, \bar{Y}^C -(\bar{p}^Y)^C\, \bar{Y}_C
\,\Bigr ] \nn \\
& \quad
- \Bigl [\,  (\bar{\theta}^\dagger)^A\, \bar{\theta}_B - (\bar{\theta}^\dagger)_B\, \bar{\theta}^A\,
\Bigr ]
+\frac{1}{2}\, \delta^A_B\, \Bigl [ (\bar{\theta}^\dagger)^C\, \bar{\theta}_C
-(\bar{\theta}^\dagger)_C\, \bar{\theta}^C\, \Bigr ] \, .
\end{align}
They are traceless and obey the $\su(2)$ algebra.

\medskip
Finally one again computes the anti-commutator 
\be\nn \{ q^A{}_{B} , {q}^C{}_{D}\} = -\frac{i}{2}\,
\epsilon^{AC}\, \epsilon_{BD} \, \int {\rm d}\sigma e^{i\, x_-}\,
\Bigl [ (P^Z)\circ Z' + (P^Y)\circ Y' + i (\theta^{\dagger}\circ
\theta' + \eta^{\dagger}\circ\eta' ) \Bigr ] \ee 
giving rise to the level-matching condition as we had in the right
(blue) algebra. We note that the right-hand side of the above
takes the same form in barred or unbarred variables.

\subsection{The centrality of the level-matching and Hamiltonian}

In this section, we show that the level-matching generator $p_{\rm
ws}$ and the Hamiltonian do Possion-commute with all the
generators of the $\su(2|2) \oplus \su(2|2)$ algebra. The explicit
computation follows the logic of section 4.

\medskip

Since we would like to work in the limit of infinite $P_+$ we also
have to suppress the corresponding conjugate zero mode $x_-$. We
pick up a solution for the unphysical field $x_-(s)$ which obeys
the boundary condition $x_-(-\infty)=0$. It reads as
$$
x_-(s)=-\zeta \int_{-\infty}^s {\rm d}\omega~\Big(p_M x'_M
-\frac{i}{2} \str(\Sigma_+\chi\chi') \Big)\, .
$$
Using the canonical Poisson brackets it is easy to find \be
\label{br}
\begin{aligned}
&\frac{1}{\zeta}\{p_M(\s),x_-(s)\}=\delta(\s-s)p_M(\s)-p'_M(\s)\epsilon(s-\s)\, , \\
&\frac{1}{\zeta}\{x_M(\s),x_-(s)\}=-x'_M(\s)\epsilon(s-\s) \, , \\
&\frac{1}{\zeta}\{x'_M(\s),x_-(s)\}=-x''_M(\s)\epsilon(s-\s)+x'_M(\s)\delta(\s-s)\,
,
\\
&\frac{1}{\zeta}\{\chi(\s),x_-(s)\}=\frac{1}{2}\delta(\s-s)\chi(\s)-\chi'(\s)\epsilon(s-\s)\,
. \end{aligned} \ee Here $\epsilon(s)$ is the standard step
function

\bea \epsilon(s)=\left\{
\begin{array}{ll} 1, ~&~ s\geq 0\, , \\ 0, ~&~
s<0
\end{array}  \right.\, ,
\eea which satisfies the condition $\epsilon(s)+\epsilon(-s)=1$.
The reader can easily verify the validity of these formulae by
considering, e.g., the Jacobi identity.
%$$
%\{p(x),\{x(y),x_-(s)\}\}+\{x_-(s),\{p(x),x(y)\}\}+\{x(y),\{x_-(s),p(x)\}\}=0\,
%.
%$$
%Using the Poisson brackets above the l.h.s. of the expression
%above can be written as
%\bea\nonumber
%\pa_y\delta(y-x)\epsilon(s-y)-\delta(x-s)\delta(y-x)+\pa_x\delta(y-x)\epsilon(s-x)
%\eea The last term here can be cast into the form \bea
%&&\pa_x\delta(y-x)\epsilon(s-x)=-\pa_y\delta(y-x)\epsilon(s-x)=-\pa_y(\delta(y-x)\epsilon(s-x))=\\
%\nonumber &&~~~~~~~~~
%-\pa_y(\delta(y-x)\epsilon(s-y))=-\pa_y\delta(y-x)\epsilon(s-y)+\delta(y-x)\delta(s-y)\,
%. \eea Thus, the Jacobi identity is perfactly satisfied.

\medskip
First, using these formulae one can check that the supercharges
commute with the level-matching generator. Introducing the
level-matching generator
$$
p_{\rm ws}=-\zeta\int_{-\infty}^\infty {\rm d}\omega~ \Big(
p_Mx'_M-\frac{i}{2} \str(\Sigma_+\chi\chi') \Big)\,
$$
it is easy to see that
$$
\frac{1}{\zeta}\{p_{\rm ws},x_M(s)\}=x'_M(s)\, ,
~~~~\frac{1}{\zeta}\{p_{\rm ws},p_M(s)\}=p'_M(s)\, ,
~~~~~~\frac{1}{\zeta}\{p_{\rm ws},\chi(s)\}=\chi'(s)
$$
and, therefore,
$$
\frac{1}{\zeta}\{ p_{\rm ws},x_-(s) \}=-\zeta\big(
p_Mx'_M-\frac{i}{2} \str(\Sigma_+\chi\chi') \big)(s)=x'_-(s)\, .
$$
Thus,
$$
\frac{1}{\zeta}\{p_{\rm ws},{\rm Q}\}=\int \pa_{s}{\rm Q}=0\,
$$
provided all the fields rapidly decrease at infinity.

\medskip

Note that $x_-(s)$ is quadratic in fermions, while we are
interested in the contribution to the Poisson bracket of $\rm H$
and $\rm Q$ which is linear in fermions. This observation implies
that computing the Poisson bracket of the Hamiltonian with $e^{i\a
x_-(s)}$ it is enough to use instead of the full $\rm H$ the
quadratic bosonic Hamiltonian with the density ${\cal H}_2^b(\s)$.
Using the basic Poisson brackets it is easy to find \bea \nonumber
\frac{1}{\zeta}\{{\cal
H}_2^b(\s),x_-(s)\}&=&(p_M^2+x'^2_M)\delta(\s-s) -\pa_{\s} {\cal
H}_2^b~\epsilon(s-\s)\, . \eea Finally, to verify the centrality
of the Hamiltonian up to order $\zeta$ we have to compute \bea
\{{\rm H},{\rm Q}\}&=& \int\int{\rm d}\s{\rm d}s \Big[ \{{\cal
H}_2^b,e^{i\a x_-}\} ~\Omega_2 + \nonumber \\
&&~~~~~~~~~~ +e^{i\a x_-}\big(\{{\cal H}_2,\Omega_2\}+\zeta
\{{\cal H}_4, \Omega_2\}+\zeta \{{\cal H }_2,\Omega_4\}\big)
\Big]+\cdots \nonumber \eea Here the integrals are taken from
$-\infty$ to $+\infty$ and to simplify the notation we do not
exhibit the dependence of functions $\Omega$ on physical fields.
We have \bea \nonumber \{{\rm H},{\rm Q}\}&=&i\a\zeta\int {\rm d}s
~e^{i\a x_-}
\big(p_M^2+x'^2_M-{\cal H}_2^b\big)\Omega_2  \\
\nonumber &&~~~+\int\int{\rm d}\s{\rm d}s ~e^{i\a x_-}\big(\{{\cal
H}_2,\Omega_2\}+\zeta \{{\cal H}_4, \Omega_2\}+\zeta \{{\cal H
}_2,\Omega_4\}\big)\, . \eea The further computation is
straightforward and it uses explicit expressions for
$\Omega_{2,4}$ in terms of transverse fields. We note that the
Poisson bracket \bea \{{\cal H}_2,\Omega_2\}+\zeta \{{\cal H}_4,
\Omega_2\}+\zeta \{{\cal H }_2,\Omega_4\} \label{cbr}\eea contains
terms proportional to $\delta(\s-s)$, $\delta'(\s-s)$ and
$\delta'' (\s-s)$ which reduces the double integration to a single
one. Moreover, according to our assumptions about the orders of
perturbation theory we are working on, in the expression
(\ref{cbr}) only the terms linear in fermions should be taken into
account. This means, in particular, that in this specific
computation only the terms in ${\cal H}_4$ which are quadratic in
fermions matter. Evaluating the brackets under these assumptions
we find that up to the order $\zeta$ the integrand appears to be a
total\ derivative and therefore vanishes for fields with rapidly
decreasing boundary conditions. Thus, with our assumptions we have
verified that
$$\{{\rm H},{\rm Q}\} =0\, ,$$ i.e. the Hamiltonian commutes with
all dynamical supercharges. It is not difficult to extend this
treatment to higher orders in fermions and in $\zeta$ but it is
already clear that we will not find any anomaly because of a rigid
structure of the supersymmetry algebra: the complete Hamiltonian
will  commute with all dynamical supercharges.

%%%%%%%%%%%%%%%%%%%%%%%%%%%%%%%%%%%%%%%%%%%%%%%%%%

%%%%%%%%%%%%%%%%%%%%%%%%%%%%%%%%%%%%%%%%%%%%%%%%%

\end{document}